\documentclass[conference]{IEEEtran}
\IEEEoverridecommandlockouts
\usepackage{hyperref}
\usepackage{cite}
\usepackage{amsmath,amssymb,amsfonts}
\usepackage{graphicx}
\usepackage{textcomp}
\usepackage{xcolor}
\usepackage{booktabs}
\usepackage{thesis}
\usepackage{wrapfig}

\usepackage{amsthm}
\newtheorem{definition}{Definition}

\AtBeginDocument{%
	\mathchardef\mathcomma\mathcode`\,
	\mathcode`\,="8000
}
{\catcode`,=\active
	\gdef,{\mathcomma\discretionary{}{}{}}
}
\mathchardef\breakingcomma\mathcode`\,
{\catcode`,=\active
	\gdef,{\breakingcomma\discretionary{}{}{}}
}

\begin{document}
\title{Striking a new balance in accuracy and simplicity with the Probabilistic Inductive Miner}

\author{\IEEEauthorblockN{Dennis Brons, Roeland Scheepens}
    \IEEEauthorblockA{UiPath Process Mining\\
    Eindhoven, The Netherlands\\
    \{dennis.brons,roeland.scheepens\}@uipath.com}
    \and
    \IEEEauthorblockN{Dirk Fahland}
    \IEEEauthorblockA{Eindhoven University of Technology\\
    The Netherlands \\
    d.fahland@tue.nl}
}


%


\maketitle

\begin{abstract}
Numerous process discovery techniques exist for generating process models that describe recorded executions of business processes. The models are meant to generalize executions into human-understandable modeling patterns, notably parallelism, and enable rigorous analysis of process deviations.
However, well-defined models with parallelism returned by existing techniques are often too complex or generalize the recorded behavior too strongly to be trusted in a practical business context.
We bridge this gap by introducing the Probabilistic Inductive Miner (PIM) based on the Inductive Miner framework. PIM compares in each step the most probable operators and structures based on frequency information in the data, which results in block-structured models with significantly higher accuracy. All design choices in PIM are based on business context requirements obtained through a user study with industrial process mining experts. PIM is evaluated quantitatively and in an novel kind of empirical study comparing users' trust in discovered model structures. The evaluations show that PIM strikes a unique trade-off between model accuracy and model complexity, that is conclusively preferred by users over all state-of-the-art process discovery methods.
\end{abstract}


\section{Introduction}\label{section:introduction}

Discovering a process model from an event log is a central step in any process mining analysis in an industrial setting. The discovered model summarizes the data and aids analysts in understanding the executed process and in distinguishing main behavior and deviations.

Research in the last 20 years contributed numerous algorithms for discovering process models supporting concurrency in syntax and semantics using human-understandable modeling notations~\cite{DeWeerdt2012ALogs,Augusto2019AutomatedBenchmark}. The models have to be sound~\cite{VanderAalst2016}, have high \emph{fitness} and \emph{precision}, and be simple in \emph{structure}~\cite{DBLP:journals/ijcis/BuijsDA14,Augusto2019AutomatedBenchmark}; best state-of-the-art techniques according to a recent benchmark~\cite{Augusto2019AutomatedBenchmark} strike different trade-offs and do not meet all criteria. For instance, the Split Miner~\cite{Augusto2018AutomatedApproach} returns models with high fitness and precision that tend to have a more complex structure and may be unsound; the Inductive Miner~\cite{Leemans2014DiscoveringBehaviour} returns sound, block-structured models with high fitness and simpler structure, but low precision.

Industrial applications still favor directly-follows graphs (DFGs) which is anecdotally attributed to their simple semantic concepts. For industrial analysts, process models with formal modeling notations are harder to understand and reason about than DFGs once models reach a certain complexity~\cite{DBLP:journals/infsof/MendlingRA10}. However, the perceived simplicity of DFGs is countered by their inability to describe processes with concurrency, leading to false statistics and wrong insights regarding performance and deviating behavior~\cite{VANDERAALST2019321} while analysts require correct information.

In this paper we (1) investigate which balance of quality properties in a process model with concurrency is preferred by analysts in an industrial setting, and (2) address the problem of identifying and developing a corresponding process discovery algorithm.

\emph{Preferred quality properties.} We answered (1) through computational requirements and a user study. As process models of an event log form a pareto-front along fitness, precision, and simplicity, models with a simpler structure have lower fitness to the  data~\cite{DBLP:journals/ijcis/BuijsDA14}. A low-fitting model can be visually augmented with the ``missing'' behavior by visually overlaying deviating paths computed from alignments~\cite{Leemans2015ExploringDeviations}. However, computing alignments is expensive~\cite{DBLP:books/sp/CarmonaDSW18} hindering quick interactive exploration of data. \emph{Visual Alignments}~\cite{Bie2019VisualThesis} are an approximation of alignments for the purpose of visualization that can be computed in linear time on block-structured BPMN models with only XOR- and AND-gateways (Appendix~\ref{sec-app:visual-alignment} shows an example). As block-structured models are also easier to understand~\cite{DBLP:journals/infsof/MendlingRA10}, we identify requirement (\textbf{R0}) \emph{The discovered model must be block-structured with only XOR- and AND-gateways.}

To further answer (1), we conducted a \emph{Delphi user study} for which we prepared 9 fragments of 3 real-life event logs. For each fragment we manually created between 2 and 8 alternative block-structured process models differing in structural complexity, fitness, and precision. We asked 6 expert process mining analysts to indicate which models they prefer as representation for each fragment (and why). Preferences and reasons were consolidated in a second round, resulting in the following requirements for process discovery in an industrial context: (\textbf{R1}) \emph{The algorithm must have a parameter to control model complexity} to allow the analyst include fewer/more details at the cost of fitness/precision. (\textbf{R2}) \emph{The algorithm must produce model structures for which there is significant evidence in the data}. Specifically, parallelism, loops, and skipping of activities should only be shown when occurring so ``frequently'' that an analyst does not doubt the the algorithm's choice.

\emph{Algorithm development.} To develop an algorithm that satisfies (R0-R2), we chose the Inductive Miner framework as it ensures block-structured models. However, all existing IM algorithms, strike the wrong trade-off to satisfy (R2), specifically IMf~\cite{Leemans2014DiscoveringBehaviour} filters infrequent behaviors using a very basic, local heuristics. The IMc~\cite{Leemans2014DiscoveringLogs} algorithm instead uses a probabilistic model to infer missing behavior. We chose to combine ideas from IMf and IMc to develop a Probabilistic IM algorithm, we called \emph{PIM}, that can handle infrequent behavior.

We recall relevant preliminaries about the IM framework and IMc in Sect.~\ref{section:prelim}. We then introduce the Probabilistic Inductive Miner (PIM) in Section \ref{section:pim}. We implemented PIM in the UiPath Process Mining platform and evaluated PIM quantitatively and empirically (see Sect.~\ref{section:evaluation}): we found that PIM strikes a unique balance between fitness, precision, and model complexity: PIM models achieve high precision while sacrificing fitness, and PIM consistently returns models with lower complexity than other algorithms. By our empirical evaluation, this unique balance is preferred by users. We present our concluding remarks in Section \ref{section:conclusion}.

\section{Background} \label{section:prelim}

We recall event logs, process trees, and the IM framework and the IMf and IMc algorithms.
A \emph{trace} $\sigma \in \Sigma^*$ is a finite sequence of activity names $\Sigma$ observed for one process execution; an \textit{event log} $L \in \mathcal{B}(\Sigma^*)$ is multi-set of traces. The \emph{directly-follows graph} (DFG) $\dfg(L)$ of $L$ has nodes $\Sigma$; an edge from $a\in \Sigma$ to $b\in\Sigma$, written $a \dfg b$, iff $b$ directly follows $a$ in some trace $\langle \ldots a,b, \ldots \rangle \in L$. Correspondingly, the \emph{indirectly-follows graph} (IFG) $\dfgSind(L)$ has an edge $a \dfgSind b$ iff $b$ strictly indirectly follows $a$ in some trace $\langle \ldots a, \ldots, b, \ldots \rangle \in L$. The \emph{frequency} $|a \dfg b|$ and $|a \dfgSind b|$ are the number of occurrences of $b$ in $L$ which directly and indirectly follow some $a$, respectively, e.g., in $\langle a,a,b,c,b,b,a,b \rangle$, $|a \dfg b| = 2$ (1st, 4th $b$) and $|a \dfgSind b| = 4$ (all 4 $b$). $\mathit{Start}(L)$ and $\mathit{End}(L)$ denote the sets of first and last activities in the traces in $L$; with correspondingly defined frequencies.

\begin{figure}[t]
   \includegraphics[width=0.6\linewidth]{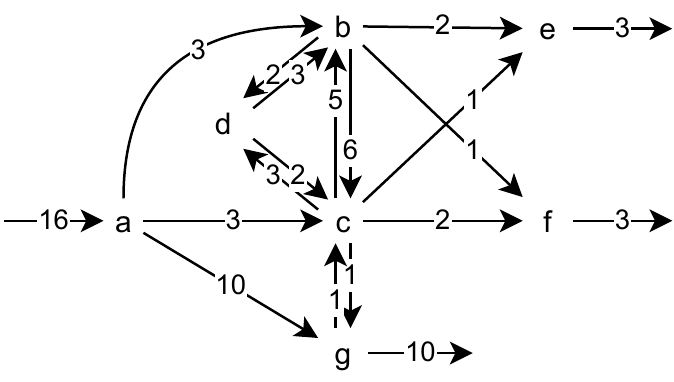}\hfill
   \includegraphics[width=0.35\linewidth]{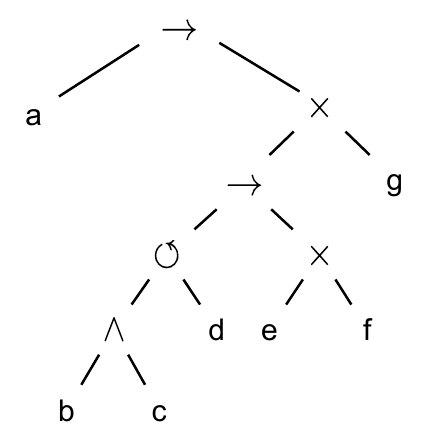}
   \caption{Directly-follows relation of $L_0$ (left) and process tree (right) from which $L_0$ was generated with some deviations.}
    \label{subfig-filter:processtree}
   \label{subfig-filter:dfg}
\end{figure}

The following log $L_0$ serves as our running example:
$L_0 = [\langle a,b,c,e \rangle, \langle a,c,b,f \rangle, \langle a,b,c,d,c,b,e \rangle^{2}  , \langle a,c,b,d,b,c,f \rangle, \langle a,c,b,d,b,c,d,b,c,f \rangle, \langle a,g \rangle^{9}, \langle a,g,c,g \rangle ]$; Fig.~\ref{subfig-filter:dfg}(left) visualizes $\dfg(L_0)$; arcs without source/target node indicate $\mathit{Start}(L_0)$ and $\mathit{End}(L_0)$.


A \emph{process tree} (PT) is an abstract representation of a sound, block-structured workflow net~\cite{Buijs2014QualitySimplicity}.
A PT is a tree where each leaf node is an activity $a \in \Sigma$ and every non-leaf node is an operator $\oplus \in \{\xor, \seq, \para, \loopnormal\}$. Each sub-tree defines a block of $\xor$ (exclusive choice), $\seq$ (sequence), $\para$ (interleaved parallelism), or $\loopnormal$ (loop) over its children; first child of $\loopnormal$ is the loop body which can be repeated after executing any of the other ``redo'' children. For example, log $L_0$ was generated from the PT in Fig.~\ref{subfig-filter:processtree}(right) with some deviations, e.g., $\langle a,g,c,g \rangle$.

The \emph{Inductive Miner} (IM)~\cite{Leemans2013DiscoveringApproach} defines a framework for recursively discovering process trees from event logs in 3 steps. (1) $\textsc{BaseCase}(L)$ checks whether $L$ has a trivial structure for which a trivial solution can be found (e.g., $L$ contains just a single activity). Otherwise, (2) $\textsc{FindCut}(L)$ identifies a identifies a \emph{cut}: a PT operator $\oplus \in \{\xor, \seq, \para, \loopnormal\}$ that best describes best the relation between $n$ partitions $\Sigma_1,...,\Sigma_n$ of the activities $\Sigma$ in $L$ (e.g., a sequence or a choice over $n$ blocks of activities). If a cut is found, (3) $\textsc{SplitLog}(L)$ splits $L$ according to the identifier operator and partitions into $L_1,\ldots,L_n$, e.g., partition the set of traces ($\xor$), sequentially split each trace ($\seq$, $\loopnormal$), or project each trace ($\para$), in a way that maximizes fitness, and recursively call $\textsc{IM}(L_1),\ldots,\textsc{IM}(L_n)$.

\emph{IM infrequent} (IMf)~\cite{Leemans2014DiscoveringBehaviour} detects an $n$-ary cut in $\dfg(L)$ by trying to \emph{top-down} partition $\Sigma$ according to an $\xor$, $\seq$, $\para$, and $\loopnormal$ operator (in this order) based on the structure of $\dfg(L)$. To detect cuts in the presence of deviations, IMf filters out relations from $\dfg(L)$ occurring relatively less often than the strongest relation at an activity. If still no cut is found, IMf returns the flower model which fits any log over $\Sigma$. But IMf's top-down approach and some design choices in filtering lead to IMf too often not finding the most likely cut in the presence of slightly ``too much'' deviating behavior. This results in low precision on real-life data~\cite{Augusto2019AutomatedBenchmark}.

In contrast, \emph{IM incomplete} (IMc)~\cite{Leemans2014DiscoveringLogs} detects a \emph{binary} cut in  $\dfg(L)$ and $\dfgSind(L)$ in a \emph{bottom-up fashion}: it computes for any pair $(a_1,a_2)$ of activities the probability $p_{\oplus} (a_1,a_2)$ that $a_1$ and $a_2$ are related by $\oplus \in \{\xor,\seq,\para,\loopnormal\}$. Then, it constructs an SMT problem to search for a partition $\Sigma = \Sigma_1 \cup \Sigma_2$ and operator $\oplus$ where the aggregated probabilities of all pairs $p_{\oplus}(a_1,a_2),a_1 \in \Sigma_1,a_2 \in \Sigma_2$ is maximal. IMc always finds the most likely cut and requires no flower model, but the $p_{\oplus}(a_1,a_2)$ are computed under the assumption of incomplete (missing) behavior in $L$ and cannot filter out deviating behavior, making it inapplicable on real-life data.

Of the other state-of-the-art algorithms~\cite{Augusto2019AutomatedBenchmark}, \emph{ETM}~\cite{Buijs2014QualitySimplicity} uses a genetic search over PTs to maximize fitness, precision, generalization and simplicity; it often finds block-structured models with better precision and fitness than IM but at very high running times. \emph{Heuristic mining techniques}~\cite{vandenBroucke2017Fodina:Technique,Augusto2017SplitLogs,DBLP:conf/icpm/AugustoDR20} determine the most likely (frequent) relations between activities through quotients over $\dfg(L)$ and $\dfgSind(L)$ and then derive split/join logic by counting succeeding/preceding activities in traces, but cannot guarantee block-structured process models; restructuring into blocks~\cite{Augusto2018AutomatedApproach} fails on most real-life logs.

In the following, we combine the ideas of IMc to detect cuts through bottom-up computation of the most likely operator between activities with the idea of quotients over $\dfg(L)$ and $\dfgSind(L)$ in logs with deviations used in heuristic miners.

\section{The Probabilistic Inductive Miner} \label{section:pim}

We now propose a new instantiation of the IM framework (cf. Sect.~\ref{section:prelim}) called \emph{Probabilistic Inductive Miner} (PIM) with the following pseudo-code.

    \begin{algorithmic}\footnotesize
    \Function{PIM}{$L, f$}
        \State $bc \gets \textproc{BaseCase}(L)$
        \If{$bc \neq \square$}
            \State{ \textbf{return} $bc$($L$) }
        \Else
            \State{$\dfg(L)^f,\dfgSind(L)^f \gets \textproc{Filtering}(\dfg(L), \dfgSind(L), f)$}
            \State{$\oplus(\Sigma_1, \Sigma_2) \gets \textproc{FindCut}(\dfg(L)^f,\dfgSind(L)^f)$}
            \State{$L_1, L_2 \gets \textproc{SplitLog}(L, \oplus(\Sigma_1, \Sigma_2))$}
            \State{$\textbf{return } \oplus (\textproc{PIM}(L_1,f), \textproc{PIM}(L_2,f)$}
        \EndIf
    \EndFunction
    \end{algorithmic}

We first discuss our new definitions of filtering the DFG (\textsc{Filtering}) in Sect.~\ref{section:alg-filtering},  \textsc{FindCut} in Sect.~\ref{section:alg-cut}, \textsc{SplitLog}  and \textsc{BaseCase} in Sect.~\ref{section:alg-split}. We apply PIM on our running example $L_0$ in Sect.~\ref{section:alg-example} before discussing some implementation details in Sect.~\ref{section:alg-rest}.

\subsection{Filtering}  \label{section:alg-filtering}

\begin{figure}[t]
    \centering
    \includegraphics[width=0.45\linewidth]{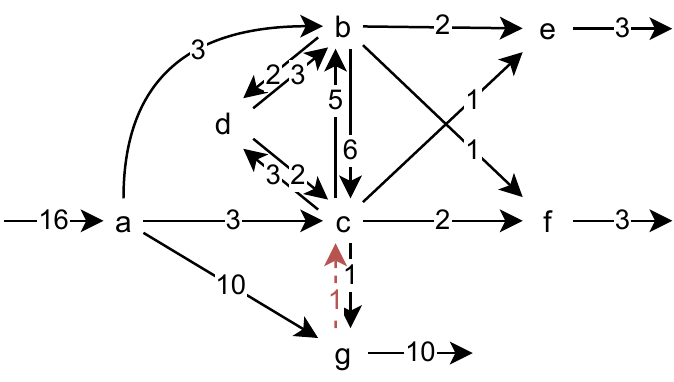}
    \includegraphics[width=0.45\linewidth]{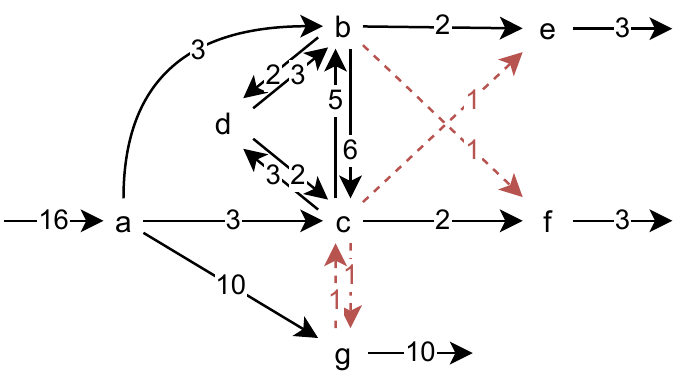}
    \caption{Filtering $\dfg(L)$ of Fig.~\ref{subfig-filter:dfg}
    by IMf with $f=0.15$ (left) and our method with $f = 97\%$ (right).}
    \label{subfig-filter:IMfStep}
    \label{subfig-filter:ourStep}
\end{figure}

The IMf algorithm~\cite{Leemans2014DiscoveringBehaviour} introduced filtering infrequent edges from $\dfg(L)$ when no cut can be found in $\dfg(L)$. IMf filters $\dfg(L)$ locally, by removing all outgoing edges $a \dfg b$ with $|a \dfg b| < f \cdot \max_{x \in \Sigma} |a \dfg x|$. This filters $\dfg(L)$ non-uniformly, as shown in Fig.~\ref{subfig-filter:IMfStep}(left) where edges $c \dfg g$ and $g \dfg c$ of same frequency are partially filtered (filtered edges in red). This impairs IMf's ability to control model complexity (BCR1) in a uniform way.

We chose to adopt percentile-based filtering for a more uniform control of model complexity. $\textproc{Filtering}(\dfg(L), \dfgSind(L), f)$ sorts all edges in $\dfg(L)$ and $\dfgSind(L)$ by their frequency, retains only the top $f\%$ of frequent (in)directly follows edges and discards all others; disconnected nodes are removed.
Filtering Fig.~\ref{subfig-filter:dfg}(left) in this way removes edges uniformly as shown in Fig.~\ref{subfig-filter:ourStep}(right). We use this method to parameterize the model complexity (BCR1).

\subsection{Cut Detection}\label{section:alg-cut}

For \textsc{FindCut}, we extend the basic idea of IMc~\cite{Leemans2014DiscoveringLogs} with principles of Heuristic Mining. First, we compute for each pair $(a,b)$ of activities a score $s_{\oplus}(a,b) \in [0,1]$ how likely it is that $a$ and $b$ are related by $\oplus \in \{\xor,\seq,\para,\loopnormal\}$ in $L$. For a given partition $\Sigma = \Sigma_1 \cup \Sigma_2$ of the activities in $L$, we then can compute an aggregated score $s_{\oplus}(\Sigma_1,\Sigma_2)$, i.e., the likelihood that the activities $\Sigma_1$ and $\Sigma_2$ form two blocks related by $\oplus$ in $L$. We then determine the cut $(\oplus, \Sigma_1,\Sigma_2)$ with $s_{\oplus}(\Sigma_1,\Sigma_2)$ maximal.

\subsubsection{Activity Relation Scores} \label{sec:alg:actrel}
\begin{figure}[t]
    \centering
    \includegraphics[trim=0cm 3cm 0cm 1.9cm, width = \linewidth]{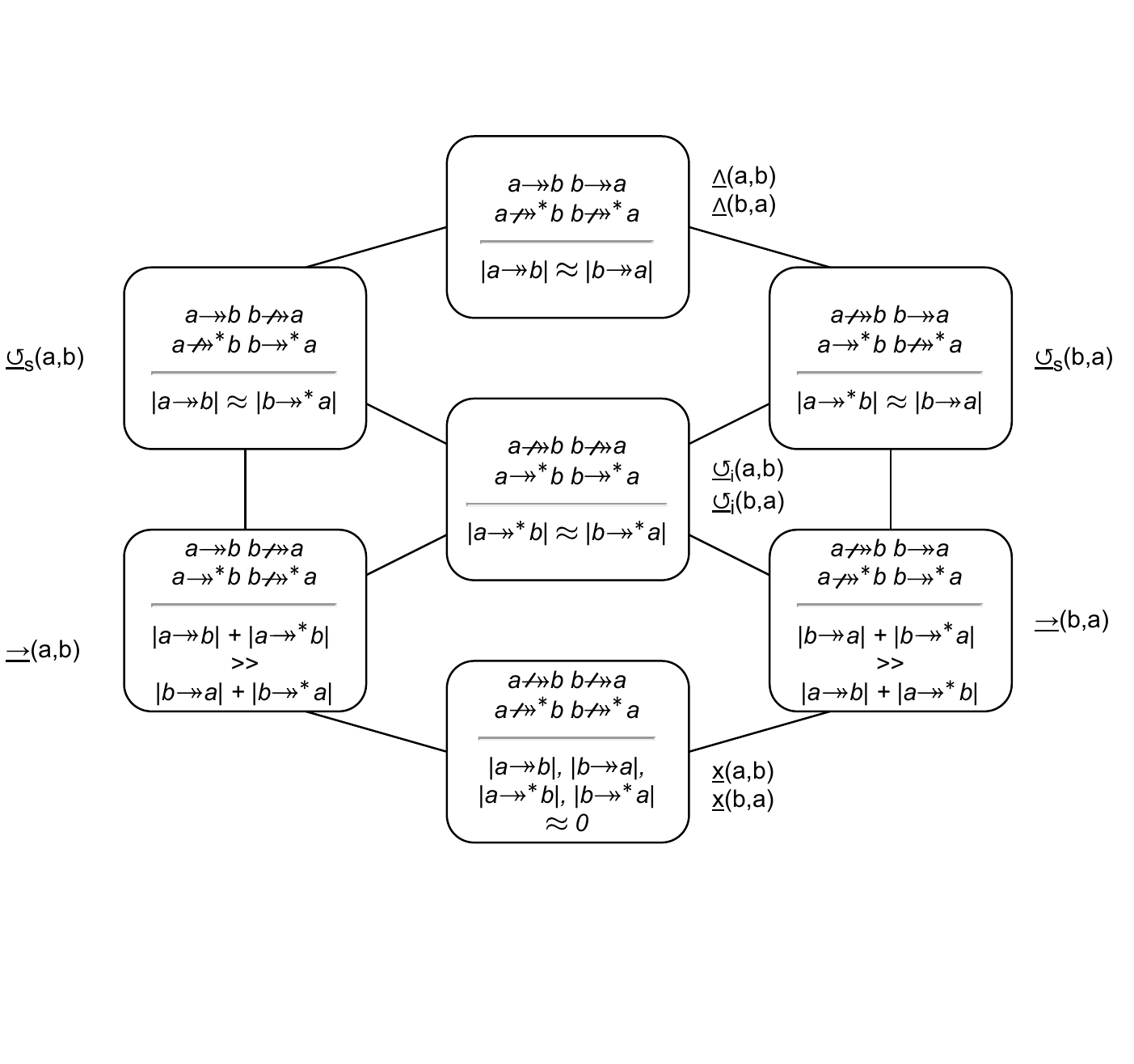}
    \caption{Activity Relation Lattice.}
    \label{pre:fig:IMLattice}
\end{figure}

The lattice in Fig.~\ref{pre:fig:IMLattice} illustrates the basic idea for determining the likelihood $s_{\oplus}(a,b)$. The IMc~\cite{Leemans2014DiscoveringLogs} introduced the mutually exclusive conditions over $\dfg$ and $\dfgSind$ shown above the horizontal line to describe when $a$ and $b$ are related by a particular operator in the absence of deviations, e.g., $a$ and $b$ are related by $\xor$ if neither follows the other (indirectly). We translated these into the fuzzy conditions over how often $a$ and $b$ (in)directly follow each other shown below the horizontal line, e.g., $a$ and $b$ are related by $\xor$ if they rarely follow each other (indirectly).


To quantify the likelihood that $\oplus$ relates $a$ and $b$ we defined the following formulas; as for IMc~\cite{Leemans2014DiscoveringLogs} for $\loopnormal$ we distinguish between $\loops$ ($a$ directly follows $b$ in a loop) and $\loopi$ ($a$ indirectly follows $b$ in a loop). In addition to $|a \dfg b|$ and $|a \dfgSind b|$ we write $|a|$ for the number of times $a$ occurs in $L$.
\begin{definition}[activity relation scores]\label{def:activity_relation_scores}
    \begin{align*}
    s_{\xor}(a,b) =  &\frac{|a| - (|a \dfg b| + |b \dfg a| + |a \dfgSind b| + |b \dfgSind a|)}{|a|}/2 + \\
    &\frac{|b| - (|a \dfg b| + |b \dfg a| + |a \dfgSind b| + |b \dfgSind a|)}{|b|} / 2
    \end{align*}  \label{alg:def:actRel:xor}
    \[
    s_{\seq}(a,b) =  \frac{|a \dfg b| + |a \dfgSind b| - (|b \dfg a| + |b \dfgSind a|)}{|a \dfg b| + |a \dfgSind b| + |b \dfg a| + |b \dfgSind a|+1}
    \]
    \[ s_{\para}(a,b) =  \min \left( \frac{|a \dfg b|}{|b \dfg a|+1} , \frac{|b \dfg a|}{|a \dfg b|+1} \right)\]
    \[ s_{\loops}(a,b) =  \min \left( \frac{|a \dfg b|}{|b \dfgSind a|+1} , \frac{|b \dfgSind a|}{|a \dfg b|+1} \right)\]
    \[ s_{\loopi}(a,b) =  \min \left( \frac{|a \dfgSind b|}{|b \dfgSind a|+1} , \frac{|b \dfgSind a|}{|a \dfgSind b|+1} \right)\]
\end{definition}
In contrast to IMc~\cite{Leemans2014DiscoveringLogs}, the scores are \emph{not} probabilities with $\sum_{\oplus} s_{\oplus}(a,b) = 1$ but heuristics measuring which of the fuzzy conditions shown in Fig.~\ref{pre:fig:IMLattice} is most likely to hold.

For $\xor$, we compare how often $a$ and $b$ occur alone, i.e., $|a|$ and $|b|$ with how often they occur together, i.e., $|a \dfg b| + |b \dfg a| + |a \dfgSind b| + |b \dfgSind a|)$. If either only $a$ or $b$ occurs in a trace, then $|a \dfg b| = \ldots = |b \dfgSind a| = 0$, and $s_{\xor}(a,b) = 1$; otherwise the second term grows and $s_{\xor}(a,b)$ approaches $0$. In $L_0$, $s_{\xor}(b,g) = 1$.

For $\seq$, we adopt the directional dependency heuristic of the Structured Heuristic Miner~\cite{Augusto2018AutomatedApproach}. $a$ is before $b$ in a sequence if the evidence for ``a precedes b'' $(|a \dfg b| + |a \dfgSind b|)$ significantly exceeds the evidence for ``b precedes a'' $(|b \dfg a| + |b \dfgSind a|)$. To keep our scores within the 0 to 1 range, we normalize the difference and truncate negative values to 0. For any activities $a$ and $b$, $s_{\seq}(a,b) = -s_{\seq}(b,a)$; $s_{\seq}(a,b)$ is maximal if $|b \dfg a| = |b \dfgSind a| = 0$. In $L_0$, $s_{\seq}(a,g) = .92$.

For $\para$, we adopt the heuristics of IMc~\cite{Leemans2014DiscoveringLogs} and Heuristic Miner~\cite{Weijters2006ProcessAlgorithm}. The more equal $|a \dfg b|$ and $|b \dfg a|$ are, the more evidence is in $L$ for a parallel relation, and the higher is $s_{\para}(a,b)$. We take the minimum of $\frac{|a \dfg b|}{|b \dfg a|+1}$ and $\frac{|b \dfg a|}{|a \dfg b|+1}$ to ensure $s_{\para}(a,b)$ lies between 0 and 1. In $L_0$, $s_{\para}(b,c) = \min ( 5 / 7, 6 / 6) = .71$.

For $\loopnormal$, we compute two scores to distinguish indirect loop relation ($s_{\loopi}(a,b)$) from entering/exiting the redo part of the loop ($s_{\loops}(a,b)$). If $a$ is in the loop body and $a \dfg b$ enters the redo part of a loop ($b$ is a redo activity), then we see $|a \dfg b|$ (entering the redo) as often as $|b \dfgSind a|$ (returning from the redo to the body), i.e., their quotient in $s_{\loops}(a,b)$ is close to 1. If $a \dfg b$ exits the redo part, then the converse $s_{\loops}(b,a)$ will be close to 1 (see Fig.~\ref{pre:fig:IMLattice}). An indirect loop relation $\loopi(a,b)$ is likely when $a$ and $b$ indirectly follow $(|a \dfgSind b|)$ and precede $(|b \dfgSind a|)$ with similar frequency.

The activity scores have the property that if $L$ is directly-follows complete and $\oplus$ (mostly) holds between $a$ and $b$ in $L$, then $s_{\oplus}(a,b) > s_{\otimes}(a,b)$ for all operators $\otimes \neq \oplus$. If $\xor$ holds, then $0 = |a \dfg b| = \ldots = |b \dfgSind a|$ and $s_{\otimes}(a,b) = 0$ for $\otimes \neq \xor$. If $\seq$ holds, then $|b \dfg a| = |b \dfgSind a| = 0$ thus $s_{\otimes}(a,b) = 0$ for $\otimes \in \{\para, \loopi, \loops \}$; also $|a \dfg b| + |a \dfgSind b| = |b|$ (c.f. Sect.~\ref{section:prelim}) thus $s_{\xor}(a,b) < .5$. If $\para$ holds, then $|a \dfg b| \approx |b \dfg a|$ thus $s_{\xor}(a,b) \approx 0$ and $s_{\seq}(a,b) \approx 0$; correspondingly $\loopi$ and $\loops$ exclude $\xor$ and $\seq$. Distinguishing $\para$ from $\loopi$ and $\loops$ requires counting repetitions~\cite{Leemans2019Information-preservingMining} which we do when aggregating activity scores.


\subsubsection{Aggregated Scores for Cuts} We now aggregate the activity relation scores $s_{\oplus}(a,b)$ to sets $s_{\oplus}(\Sigma_1,\Sigma_2)$ of activities $\Sigma = \Sigma_1 \cup \Sigma_2$ so that we can search for cuts $(\oplus,\Sigma_1,\Sigma_2)$ with $s_{\oplus}(\Sigma_1,\Sigma_2)$ being maximal. We write $S(\oplus,R) = [ s_{\oplus}(a,b) \mid (a,b) \in R ]$ for the bag of scores of activity pairs $R$.

For $L$ having no deviations, IMc defined
$s_{\oplus}(\Sigma_1,\Sigma_2)$ as the \emph{average} $\mu(S(c)) = \frac{\sum_{s \in S(c)} s}{|S(c)|}$ of all activity relation scores $S(c) = S(\oplus,\Sigma_1\times\Sigma_2)$ in cut $c$, with a special case for $\oplus = \loopnormal$~\cite{Leemans2014DiscoveringBehaviour}. However, deviations in $L$ may give a few pairs $(a,b)$ a biased score $s_{\otimes}(a,b) > s_{\oplus}(a,b)$ which could lead to a ``wrong'' average $\mu(S(\otimes,\Sigma_1' \times \Sigma_2')) > \mu(S(\oplus,\Sigma_1\times\Sigma_2))$ although $s_{\otimes}(x,y) < s_{\oplus}(x,y)$ for most pairs $(x,y)$. We therefore introduce a \emph{correction term or factor} to obtain a lower score $s_{\otimes}(\Sigma_1',\Sigma_2') < \mu(S(\otimes,\Sigma_1',\Sigma_2'))$ when we see evidence of such wrong bias. We first present the formal definitions and provide a full example in Sect.~\ref{alg:sec:example}.

For $\oplus \in \{\xor,\seq\}$, we know from Sect.~\ref{sec:alg:actrel} that $s_{\oplus}(a,b)$ is low iff $s_{\otimes}(a,b), \oplus \neq \otimes$ is high ($a$ and $b$ related by $\otimes$ and not by $\oplus$). Thus, a high average over $s_{\oplus}(a,b), (a,b) \in \Sigma_1\times\Sigma_2$ is falsely biased towards $\oplus$ if some pairs $(a,b)$ have a significantly lower score $s_{\oplus}(a,b)$ than other pairs. We therefore use as correction term for $\xor$ and $\seq$ the \emph{standard deviation} $\sigma(S)$ over a set $S$ of activity relation scores.
\begin{definition}[aggregate score for $\xor$, $\seq$] \label{def:alg:modAccScore:XS}
    Let $\Sigma_1 \cup \Sigma_2 = \Sigma$; $\oplus \in \{ \xor, \seq \}$. $c = (\oplus, \Sigma_1, \Sigma_2)$ is an $\oplus$-\emph{cut} with score $s_\oplus(\Sigma_1, \Sigma_2) = \mu(S(c)) - \sigma(S(c))$.
\end{definition}
For $\oplus \in \{\para,\loopnormal\}$, we know from Sect.~\ref{sec:alg:actrel} and Def.~\ref{def:activity_relation_scores} that (1) if $s_{\otimes}(a,b),\otimes \in \{\xor,\seq\}$ is low then $(a,b)$ can be in $\para$ or $\loopnormal$ relation, and vice versa, and that (2) we cannot reliably distinguish parallel behavior from loop behavior using only binary activity relations.

Distinguishing $\para$ from $\loopnormal$ requires explicitly counting repetitions~\cite{Leemans2019Information-preservingMining}, which we achieve as follows.
We write $|L|$ for the number of traces and $||L||$ for the total number of events in log $L$. If $L$ has no repetition (no loop) then each trace in $L$ contain each $a \in \Sigma$ at most once and $||L|| \leq |L| \cdot |\Sigma|$. Thus, $r(L) = \frac{|L|}{\left( \frac{||L||}{|\Sigma|}\right)} < 1$ if $L$ has a loop. We thus can use $r(L)$ as correction factor to reduce the $\para$ score (presence of loops), which we bound to $[0;1]$ by $\min (r(L), 1)$.
\begin{definition}[aggregate score for $\para$] \label{def:alg:modAccScore:P}
    Let $\Sigma_1 \cup \Sigma_2 = \Sigma$. Then $c = (\para, \Sigma_1, \Sigma_2)$ is a $\para$-cut with \emph{score} $s_{\para}(\Sigma_1, \Sigma_2) = \mu(S(c)) \cdot \min (r(L), 1)$. \label{def:alg:modAccScore:P}
\end{definition}
As $r(L) \approx 1$ or $r(L) \geq 1$ only indicates absence of $\loopnormal$ but not presence of $\para$, we have to use the inverse $(1- \min (r(L), 1))$ to boost a $\loopnormal$-score (reinforcing presence of loops). IMc~\cite{Leemans2014DiscoveringLogs} computes the score for $\loopnormal$ as average over $\loopi$ and $\loops$ as follows. In a loop, the redo part $\Sigma_2$ is entered from/exited to body $\Sigma_1$ at $S_2,E_2 \subseteq \Sigma_2$, respectively. Pairs $(a,b)$ directly entering/exiting the redo part are scored using $\loops$, while all other (indirect) pairs are scored using $\loopi$ (see Fig.~\ref{pre:fig:IMLattice}).
%
\begin{definition}[aggregated score for $\loopnormal$] \label{def:alg:modAccScore:L}
    Let $\Sigma_1 \cup \Sigma_2 = \Sigma$ and $S_2, E_2 \subseteq \Sigma_2$. Then $c = (\loopnormal, \Sigma_1, \Sigma_2, S_2, E_2)$ is a $\loopnormal$-\emph{cut} defining sets $\mathit{enter} = \mathit{End}(L) \times S_2$, $\mathit{exit} = E_2 \times \mathit{Start}(L)$, and $\mathit{indirect} = (\Sigma_1 \times \Sigma_2) \setminus (\mathit{enter} \cup \mathit{exit})$.

    The scores for $c$ are $S(c) = S(\loops,\mathit{enter} \cup \mathit{exit}) \cup S(\loopi,\mathit{indirect})$.
    The \emph{aggregated score} for $c$ is $s_{\loopnormal}(\Sigma_1, \Sigma_2, S_2, E_2) = \mu(S(c)) + (\mu(S(c)) \cdot (1 - \min(r(L),1)))$.
\end{definition}
The correction term $+ (\mu(S(c)) \cdot (1 - \min(r(L),1))$ boosts the loop score relative to the inverse $(1 - \min(r(L),1))$ which is high when $r(L) \ll 1$, i.e., $L$ shows many repetitions.

\subsubsection{Cut finding}\label{alg:sec:cut-finding} A na\"{i}ve method for \textsc{FindCut} computes all activity relation scores $s_{\oplus}(a,b)$ for all $(a,b) \in \Sigma \times \Sigma$ and then exhaustively searches for a partition $\Sigma_1 \cup \Sigma_2$ with $s_{\oplus}(\Sigma_1,\Sigma_2)$ or $s_{\loopnormal}(\Sigma_1,\Sigma_2,E_2,S_2), E_2,S_2\subseteq\Sigma_2$ being maximal, which costs $O(|\Sigma|^2)+O(4^{|\Sigma|})$. Heuristics in $\dfg$ and $\dfgSind$ allow to significantly prune the search space for $\Sigma_1,\Sigma_2,E_2,S_2$ but are omitted for space limitations. Note that this method always returns a cut (possibly with a low score) while IMf may not find a cut and resort to a flower model~\cite{Leemans2014DiscoveringBehaviour}.

\subsection{Log Splitting, Recursion, Base Cases, Skipping} \label{alg:sec:logsplitting}\label{section:alg-split}
Following the IM framework, we next split the log $L$ according to the found cut $c = (\oplus,\Sigma_1,\Sigma_2)$ into $L_1$ and $L_2$. As $c$ may not completely fit $L$ due to deviations, we base \textsc{SplitLog} on IMf's~\cite{Leemans2014DiscoveringBehaviour} method which filters from $L_1$ and $L_2$ events that deviate from $(\oplus,\Sigma_1,\Sigma_2)$. Filtering may result in empty traces $\langle \rangle\in L_i$, denoting that the behavior in $L_i$ can also be skipped. Then \textsc{Pim} is invoked on $L_1$ and $L_2$, returning a left and right sub-tree under $\oplus$ (see start of Sect.~\ref{section:pim}).

We use the IM base cases, with some extensions regarding the handling of empty behavior in a (sub)log.
\begin{itemize}[noitemsep, nolistsep, leftmargin=*]
    \item \textproc{Single activity} ($|\Sigma| = 1$). When the sublog $L$ contains only a single activity, that activity is returned as a leaf node of the process tree.
    \item \textproc{No Activities} ($|\Sigma| = 0$). If the sublog $L$ does not contain any activities, a silent activity ($\tau$) is returned as leaf node of the process tree.
    \item \textproc{Skip Sublog} When the sublog contains $n$ empty traces $\langle\rangle^n \in L$, a skip sub-tree $\xor(\tau,\textsc{Pim}(L \setminus \{ \langle \rangle \} ))$ is returned as explained next.
\end{itemize}
IMf's \textsc{BaseCase} always filters $\langle \rangle^n$ from $L$ and returns  $\xor(\tau,\textsc{IMf}(L \setminus \{ \langle \rangle \} ))$ if $n \geq |L|\cdot f$ wrt. threshold $f$. This either forgets that some empty behavior has been seen or results in many $\tau$-skips across all subtrees which lowers precision~\cite{Augusto2019AutomatedBenchmark}. We delay generating  $\xor(\tau,\ldots)$ as follows. If $n \leq .5\cdot|L|$, we keep all empty traces in $L$ but ignore them when computing $\dfg$, $\dfgSind$, and $r(L)$. Upon log splitting, $L_1$ and $L_2$ both get all $\langle\rangle^n$ empty traces and filtering may add further ones. Only when we accumulated $n > .5\cdot|L|$ empty traces, i.e., the majority of the behavior in $L$ is skipping, \textproc{Skip Sublog} introduces the $\xor(\tau,\ldots)$ which addresses (R2).

\subsection{Example}\label{section:alg-example}
\begin{wrapfigure}[8]{l}{.45\linewidth}\centering
\includegraphics[width=\linewidth, valign=t]{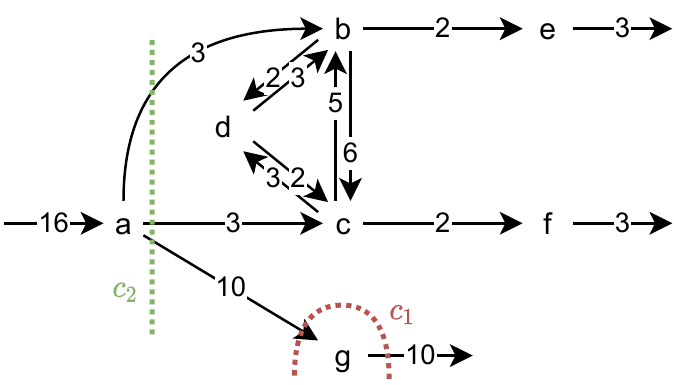}
\caption{Cut $c_1$ vs $c_2$}\label{subfig-mine:c1c2}
\end{wrapfigure}
Consider $L_0$ with filtered DFG in Fig.~\ref{subfig-filter:ourStep}(right) and cuts $c_1 = (\xor, \{a,b,c,d,e,f\}, \{g\})$ and $c_2 = (\seq, \{a\}, \{b,c,d,e,f,g\})$ in Fig.~\ref{subfig-mine:c1c2}. The mean scores (as in IMc) are $\mu(c_1) = .86 > \mu(c_2) = .79$, because $g$ has strong $\xor$ relations to $b,c,d,e,f$. However, $s_{\xor}(a,g) = .16 \ll 1$ shows that some other relation holds between $a$ and $g$ and $\mu(c_{1})$ is falsely biased towards $\xor$; this shows in $\sigma(c_{1}) = .29$. In contrast, $s_{\seq}(a,y)$ is high for \emph{all} $y \in \{b,c,d,e,f,g\}$ due to $a \dfg y$ or $a \dfgSind y$ (no other activity is likely), and $\sigma(c_{2}) = .13$. Thus, $s_{\xor}(\{a,b,c,d,e,f\}, \{g\}) = .54$ and $s_{\seq}(\{a\}, \{b,c,d,e,f,g\}) = .67$, making $c_{2}$ the more likely cut.

\begin{wrapfigure}[8]{l}{.3\linewidth}\centering
\includegraphics[width=\linewidth, valign=t]{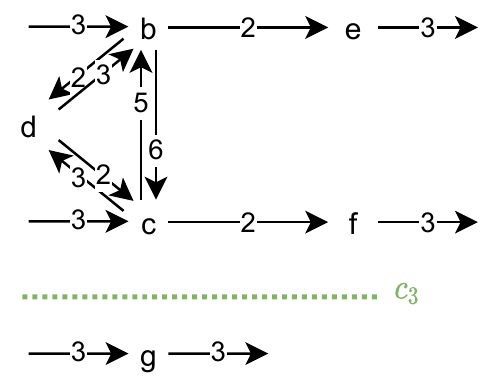}
\caption{Cut $c_3$}\label{subfig-mine:c3}
\end{wrapfigure}
After log splitting and applying \textsc{BaseCase} on $L_1 = [ \langle a \rangle^{16} ]$, PIM finds $c_3 = (\xor, \{b,c,d,e,f\}, \{g\})$ on $\dfg$ and $\dfgSind$ of $L_2$ in Fig.~\ref{subfig-mine:c3}. Log splitting filters $c$ from $L_3 = [\langle g \rangle,\langle g,c,g \rangle$ \textsc{BaseCase} returns $g$. On $L_4 = [\langle b,c,e \rangle, \langle c,b,f \rangle, \langle b,c,d,c,b,e \rangle^{2}, \langle c,b,d,b,c,f \rangle, \langle c,b,d,b,c,d,b,c,f \rangle ]$.

\begin{figure}[t]
    \centering
    \subfloat[\label{subfig-mine:c4}]{%
      \includegraphics[scale=0.6, valign=t]{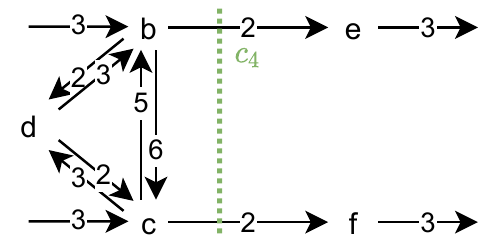}
    }\hfill
    \subfloat[\label{subfig-mine:c5}]{%
      \includegraphics[scale=0.6, valign=t]{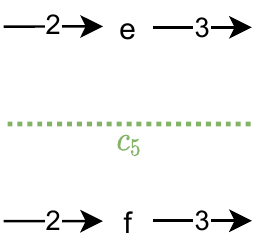}
    }\hfill
    \subfloat[\label{subfig-mine:c6c7}]{%
      \includegraphics[scale=0.6, valign=t]{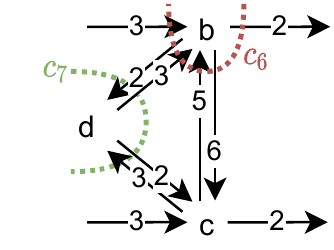}
    }\hfill
    \subfloat[\label{subfig-mine:c8}]{%
      \includegraphics[scale=0.6, valign=t]{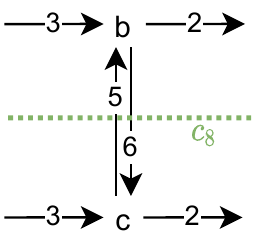}
    }
    \caption{Cuts $c_4$-$c_8$ show effect of correction factor in  Def.\ref{def:alg:modAccScore:XS}-\ref{def:alg:modAccScore:L}.}
    \label{fig:filter}
\end{figure}

Subsequently, PIM calculates and chooses cut $c_4 = (\seq, \{b,c,d\}, \{e,f\})$, see Fig.~\ref{subfig-mine:c4}. For the right-hand side, PIM finds $c_5 = (\xor, \{e\}, \{f\})$ with base cases for $e$ and $f$, see Fig.~\ref{subfig-mine:c5}, respectively. In $\dfg$ and $\dfgSind$ of $L_5 = [\langle b,c \rangle, \langle c,b \rangle, \langle b,c,d,c,b \rangle^{2}, \langle c,b,d,b,c \rangle, \langle c,b,d,b,c,d,b,c \rangle ]$ of the left-hand side (Fig.~\ref{subfig-mine:c6c7}) we see the effect of correction term $r(L_5)$ for distinguishing $\para$ and $\loopnormal$.

Consider cuts $c_6 = (\para, \{b\}, \{c,d\})$ and $c_7 = (\loopnormal, \{b,c\}, \{d\})$in Fig.~\ref{subfig-mine:c6c7}. Without correction, $\mu(S(c_6)) = 1.0$ and $\mu(S(c_7)) = .79$. As $r(L_5) = .75$, the correction factors yield $s_{\para}(\{b\}, \{c,d\}) = \mu(S(c_6)) \cdot r(L_5) = .75$ and $s^m_{\loopnormal}(\{b,c\}, \{d\}) = \mu(S(c_7)) \cdot (2 - .75) = .99$, and PIM returns $c_7$. After log splitting, PIM finds $c_8 = (\para, \{b\}, \{c\})$ on $L_6 = [\langle b,c \rangle^{6}, \langle c,b \rangle^{5}] $ at the ``correct'' location (see Fig.~\ref{subfig-mine:c8}). Applying \textsc{BaseCase} then results in the process tree of Fig.~\ref{subfig-filter:processtree}.

\subsection{Complexity} \label{alg:sec:summary}\label{section:alg-rest}

The running time complexity of PIM depends on the size of the event log $L$, and the size of its alphabet $\Sigma$ as follows. Per recursion step, $\textproc{BaseCase}$ is executed once, constructing and filtering $\dfg$ and $\dfgSind$ takes $O(|\Sigma|^2)$, \textproc{FindCut} takes $O(4^{|\Sigma|})$ (see Sect.~\ref{alg:sec:cut-finding}), and \textproc{Split Log} takes $O(|L|)$, decreasing the size of the alphabet in each sublog by at least one. Thus, there are at most $\Sigma$ invocations of \textsc{Pim} resulting in running-time complexity of $O(|\Sigma| \cdot (1 + |\Sigma|^2 + 4^{|\Sigma|} + |L|)) = O(|\Sigma| \cdot (4^{|\Sigma|} + |L|))$. As discussed in Sect.~\ref{alg:sec:cut-finding}, heuristics over $\dfgind$ and $\dfgSind$ reduce the search space for \textproc{FindCut} significantly as measured in our experiments.


\section{Evaluation} \label{section:evaluation}

We implemented PIM in the UiPath Process Mining platform and compared it to the state-of-the-art on a standard benchmark~\cite{Augusto2019AutomatedBenchmark} and on additional data more similar to regular use cases encountered in industrial practice (\ref{section:evaluation:benchmark}). Additionally, we performed an empirical evaluation about end user preferences for models produced by various algorithms (\ref{section:evaluation:empirical}).

\subsection{Quantitative Evaluation}\label{section:evaluation:benchmark}
\emph{Setup.} The existing benchmark~\cite{Augusto2019AutomatedBenchmark} uses a set of non-synthetic event logs to comparatively evaluate 7 automated process discovery methods of which Evolutionary Tree Miner (ETM)\cite{Buijs2014QualitySimplicity}, IMf~\cite{Leemans2014DiscoveringBehaviour} and Split Miner (SM)~\cite{Augusto2017SplitLogs}, outperformed the other methods. We therefore compare PIM to ETM, IMf, and SM on the public benchmark event logs of~\cite{Augusto2019AutomatedBenchmark}, see Tab. \ref{tab:datasets}. Model accuracy is measured by fitness, precision, and their F-score; simplicity is measured by size, control-flow complexity (CFC), structuredness, and soundness, see~\cite{Augusto2019AutomatedBenchmark}.

However, the benchmark is not representative of industrial workloads: (1) Most event logs were prefiltered to reduce complexity~\cite{Augusto2019AutomatedBenchmark} while techniques in industrial practice face unfiltered logs, (2) industrial logs are larger, and (3) the benchmark lacks certain process types found in practice. We thus included 4 additional unfiltered datasets: 2 public datasets BPIC14 and BPIC17\textsubscript{LC} (distinguish activity life-cycles), and 2 proprietary datasets (Invoice and P2P); see Tab~\ref{tab:datasets}.

\begin{table}[t]
\centering
\begin{tabular}{  c || c c | c c }
\textit{Log} & \textit{Total} & \textit{Distinct}   & \textit{Total} & \textit{Distinct}\\

\textit{Name} & \textit{Traces} & \textit{Traces (\%)} & \textit{Events}  & \textit{Events}\\
\hline
\hline
BPIC12 & 13,087 & 33.4 & 262,200 & 36 \\
\hline
BPIC13\textsubscript{cp} & 1,487 & 12.3 & 6,660 & 7 \\
\hline
BPIC13\textsubscript{inc} & 7,554 & 20.0 & 65,533 & 13 \\
\hline
BPIC14\textsubscript{f} & 41,353 & 36.1 & 369,485 & 9 \\
\hline
BPIC15\textsubscript{1f} & 902 & 32.7 & 21,656 & 70 \\
\hline
BPIC15\textsubscript{2f} & 681 & 61.7 & 24,678 & 82 \\
\hline
BPIC15\textsubscript{3f} & 1,369 & 60.3 & 43,786 & 62 \\
\hline
BPIC15\textsubscript{4f} & 860 & 52.4 & 29,403 & 65 \\
\hline
BPIC15\textsubscript{5f} & 975 & 45.7 & 30,030 & 74 \\
\hline
BPIC17\textsubscript{f} & 21,861 & 40.1 & 714,198 & 41 \\
\hline
RTFMP & 150,370 & .2 & 561,470 & 11 \\
\hline
SEPSIS & 1,050 & 80.6 & 15,214 & 16 \\
\hline
\hline
BPIC14 & 46,616 & 48.5 & 466,737 & 39  \\
\hline
BPIC17\textsubscript{LC} & 31,509 & 53.0 & 1,202,267 & 66  \\
\hline
Invoice & 24,450 & 1.6 & 133,452 & 15 \\
\hline
P2P & 616,717 & 15.5  & 5,583,650 & 46 \\
\end{tabular}
\caption{Statistics of the public event logs, extracted from \cite{Augusto2019AutomatedBenchmark}, followed by the additional event logs.}
\label{tab:datasets}
\end{table}

We ran PIM twice: (1) bounding the na\"{i}ve \textsc{FindCut} with complexity $O(4^{|\Sigma|})$ (Sect.~\ref{alg:sec:cut-finding}) to only find cuts over the 30 most frequent activities to adhere to the 4 hour computation limit set in~\cite{Augusto2019AutomatedBenchmark} (PIM\textsubscript{30}); (2) using fast heuristics in \textsc{FindCut} for which no activity bound was required; in both cases $f=99.5\%$ to filter only the most infrequent behavior. On the 4 additional logs, we ran IMf and SM with  default parameters, but omit EMT due to its excessive running times.


\emph{Results.}
PIM and PIM\textsubscript{30} discovered sound, block-structured models on all event logs while SM returned unsound models on BPIC14 and BPIC17\textsubscript{LC}.

\begin{figure}[t]\centering
\includegraphics[width=\linewidth]{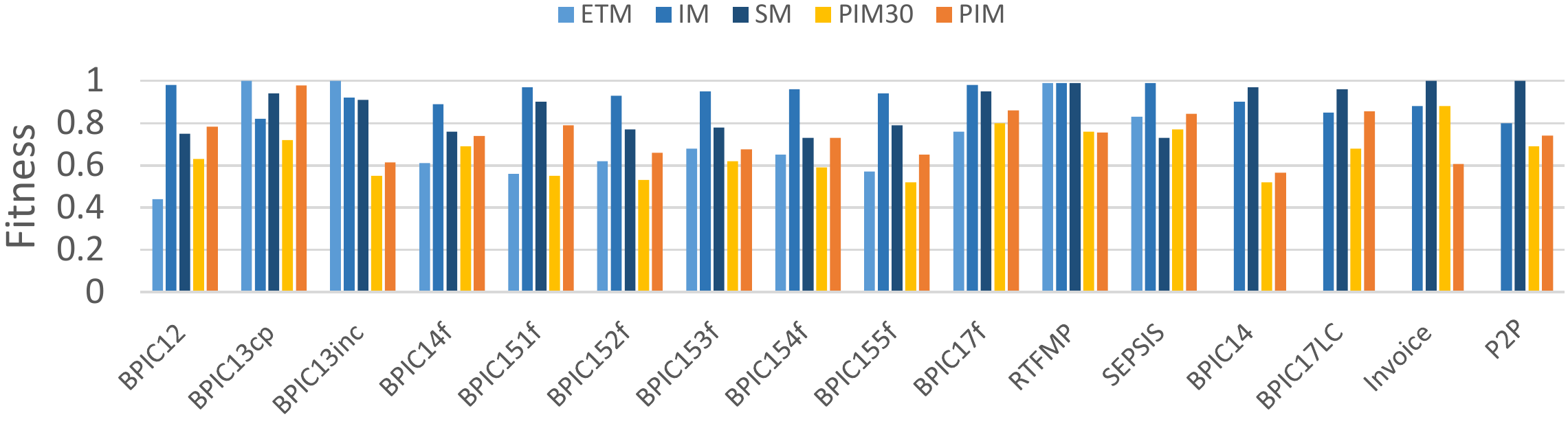}
\includegraphics[width=\linewidth]{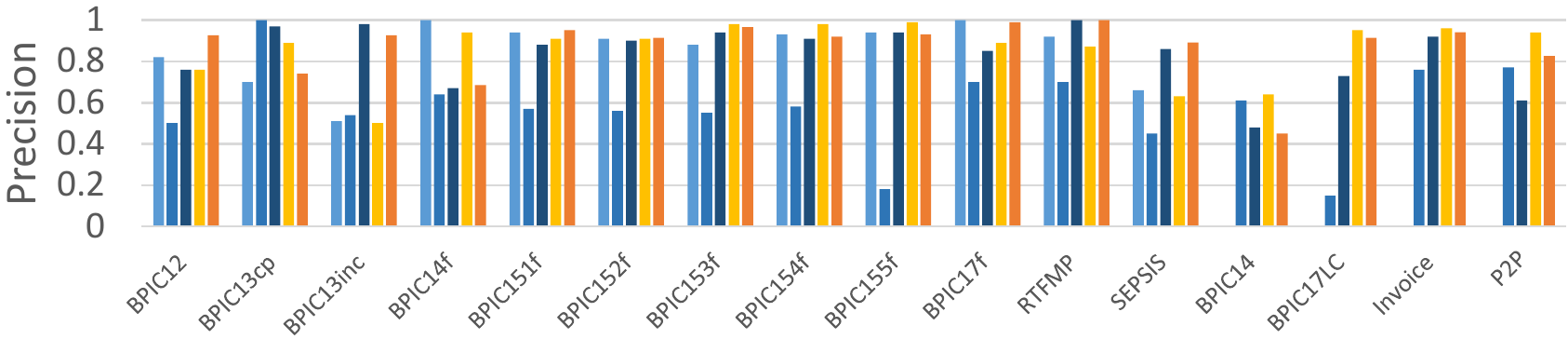}
\includegraphics[width=\linewidth]{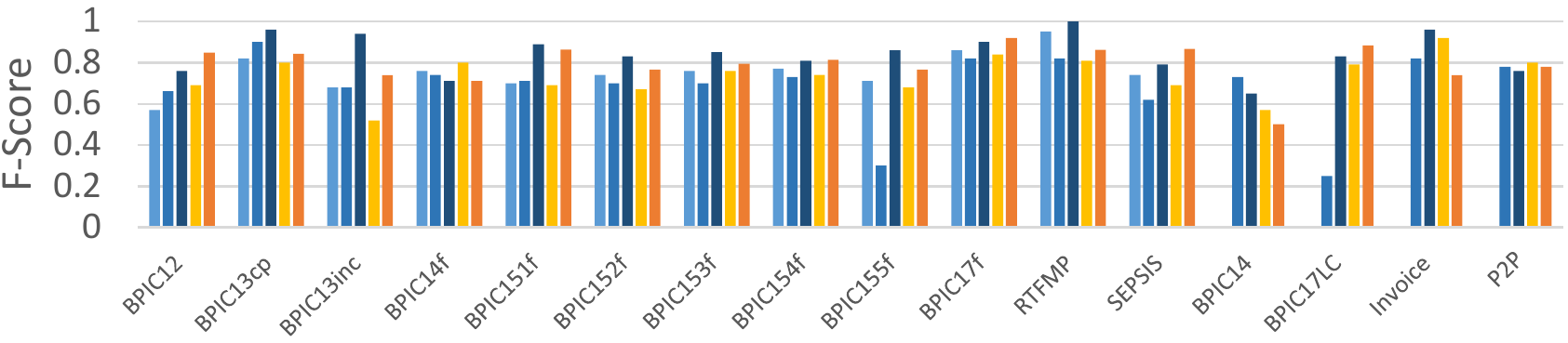}
\includegraphics[width=\linewidth]{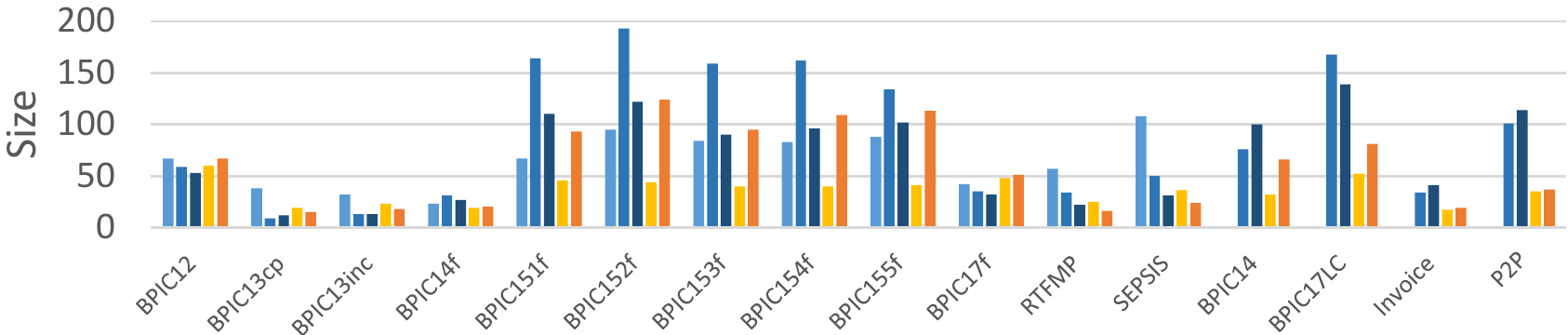}
\includegraphics[width=\linewidth]{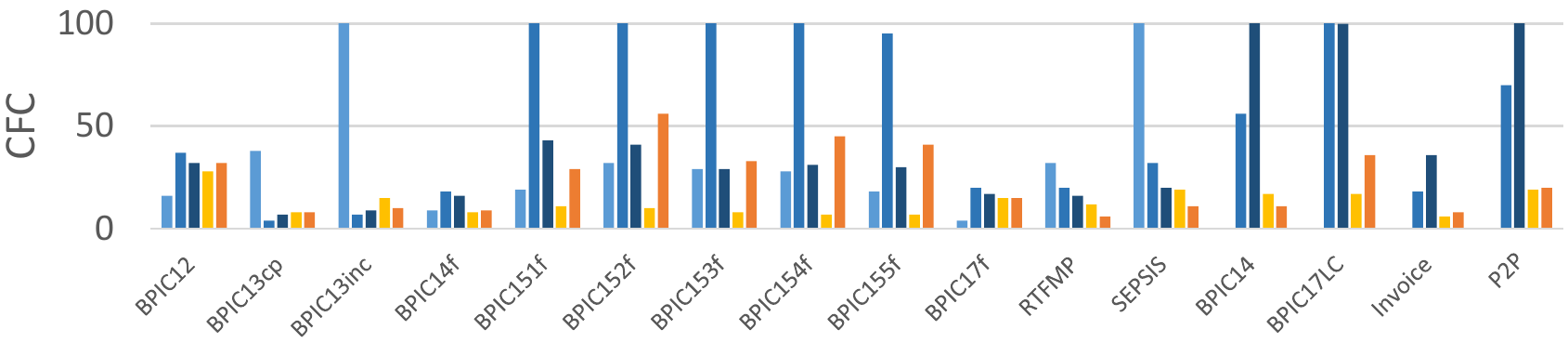}
\includegraphics[width=\linewidth]{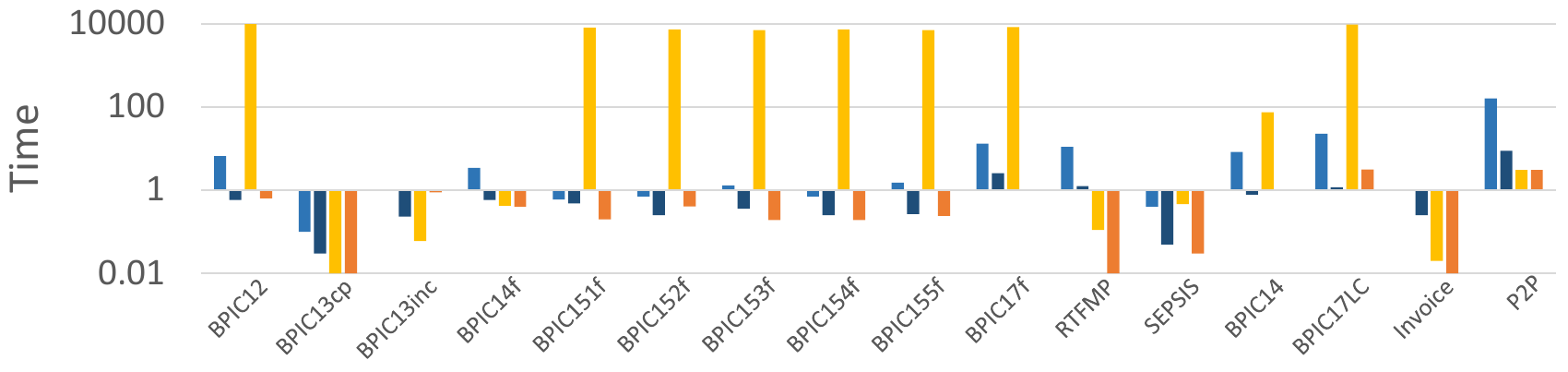}
\caption{Fitness, precision, f-score, size, CFC, and running time for ETM, IMf, SM~\cite{Augusto2019AutomatedBenchmark}, PIM\textsubscript{30}, and PIM}\label{fig:results_all}
\end{figure}

\begin{figure}[t]
    \centering
    \includegraphics[height=2.2cm]{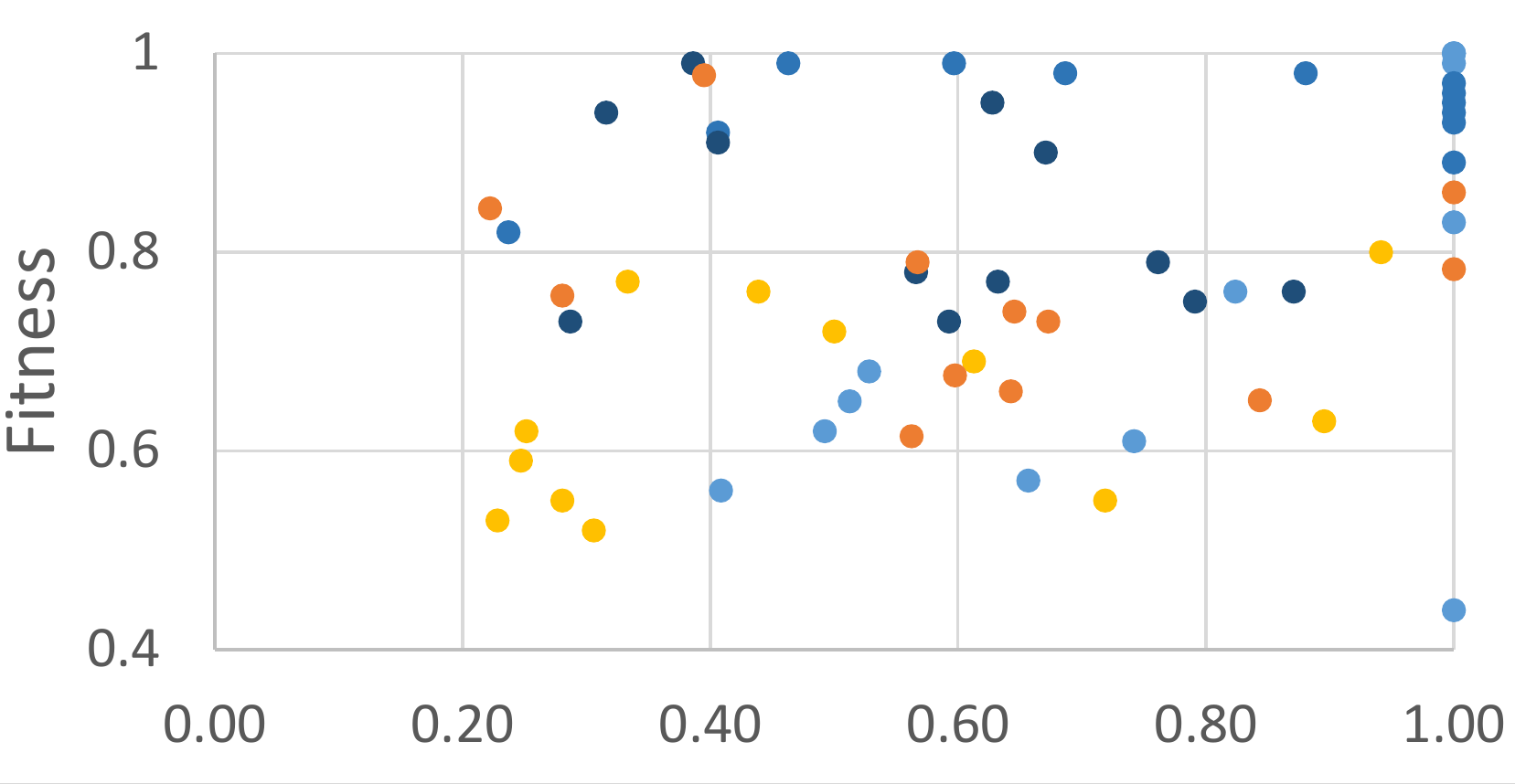}
    \includegraphics[height=2.2cm]{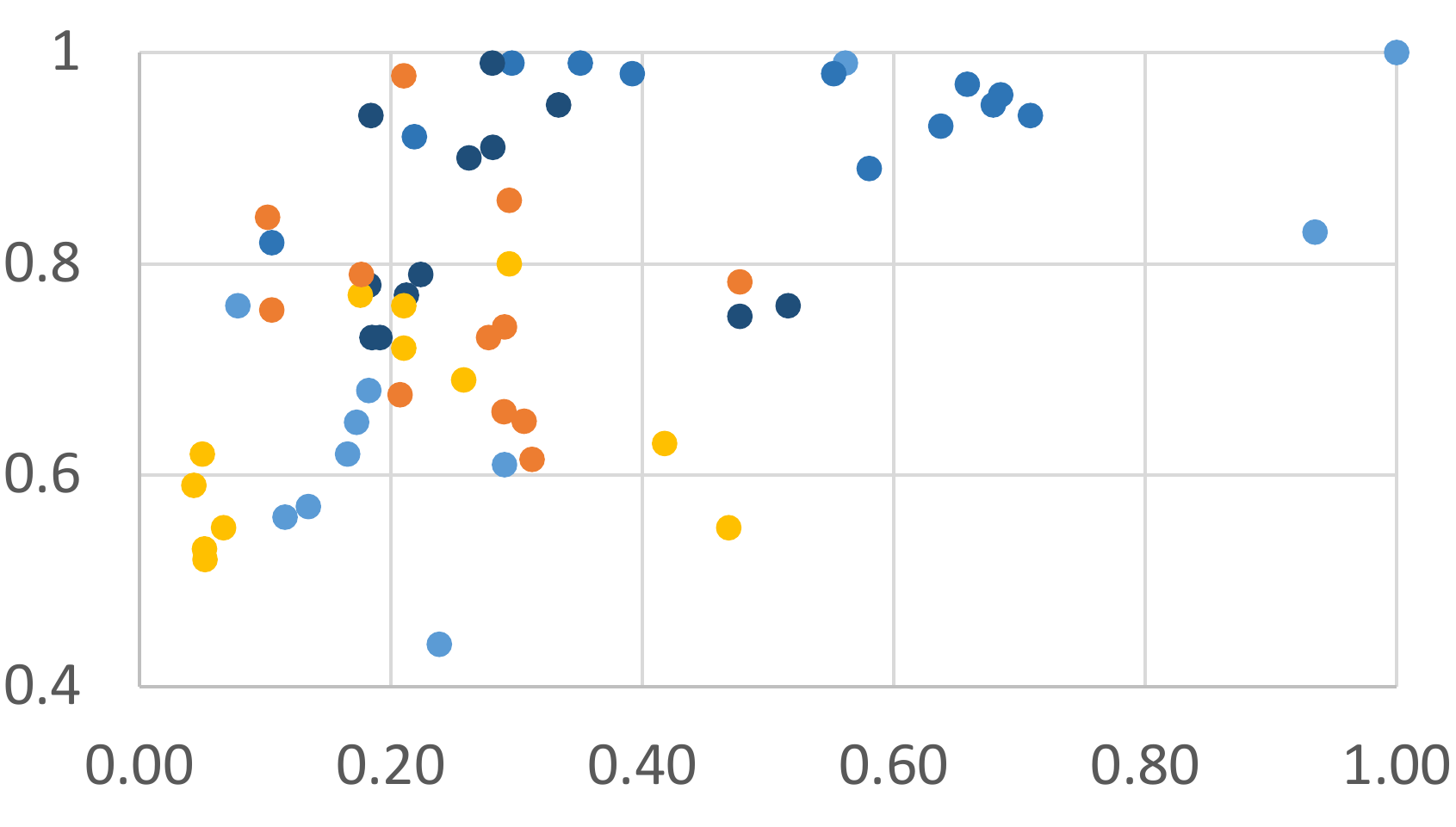}
    \includegraphics[height=2.2cm]{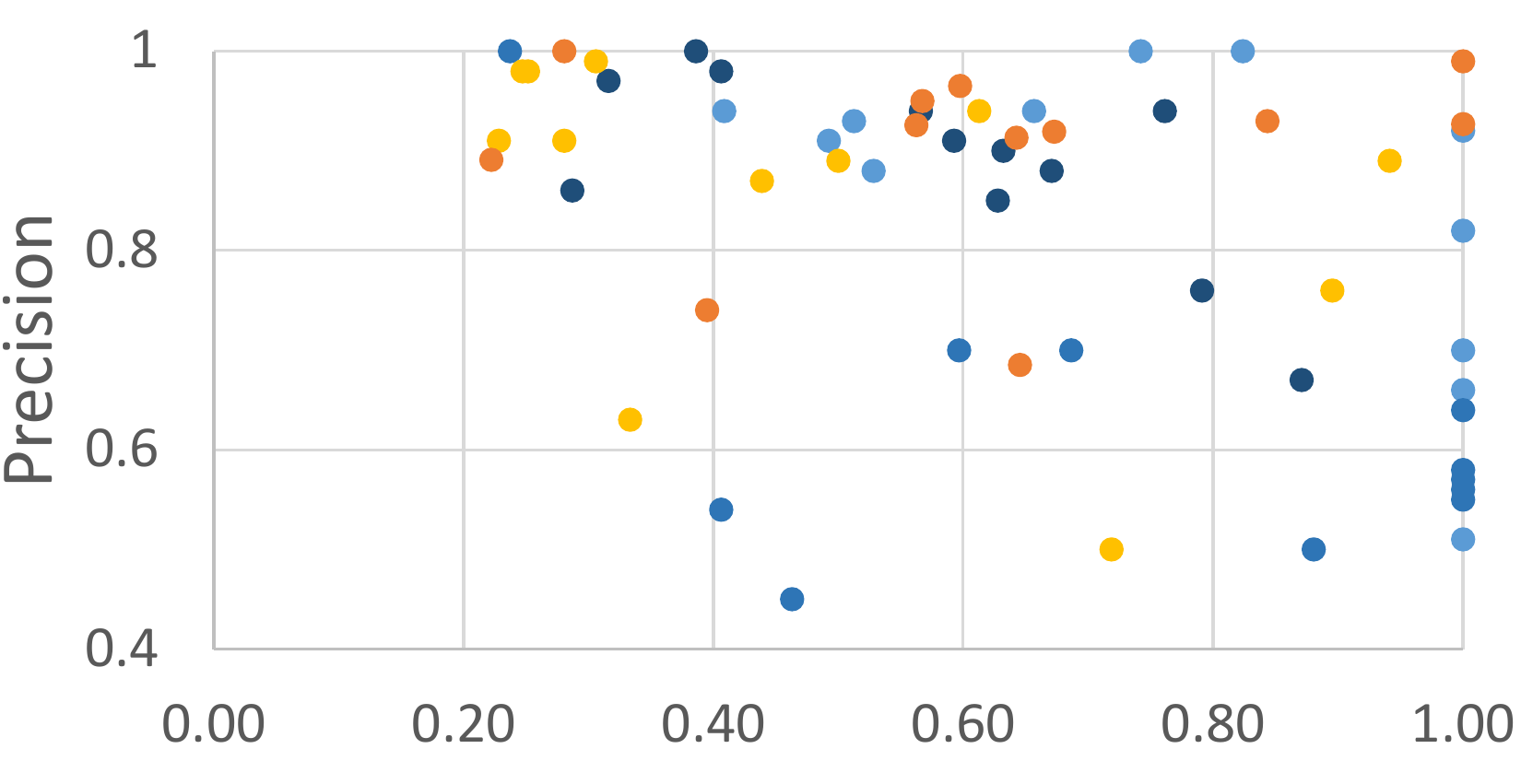}
    \includegraphics[height=2.2cm]{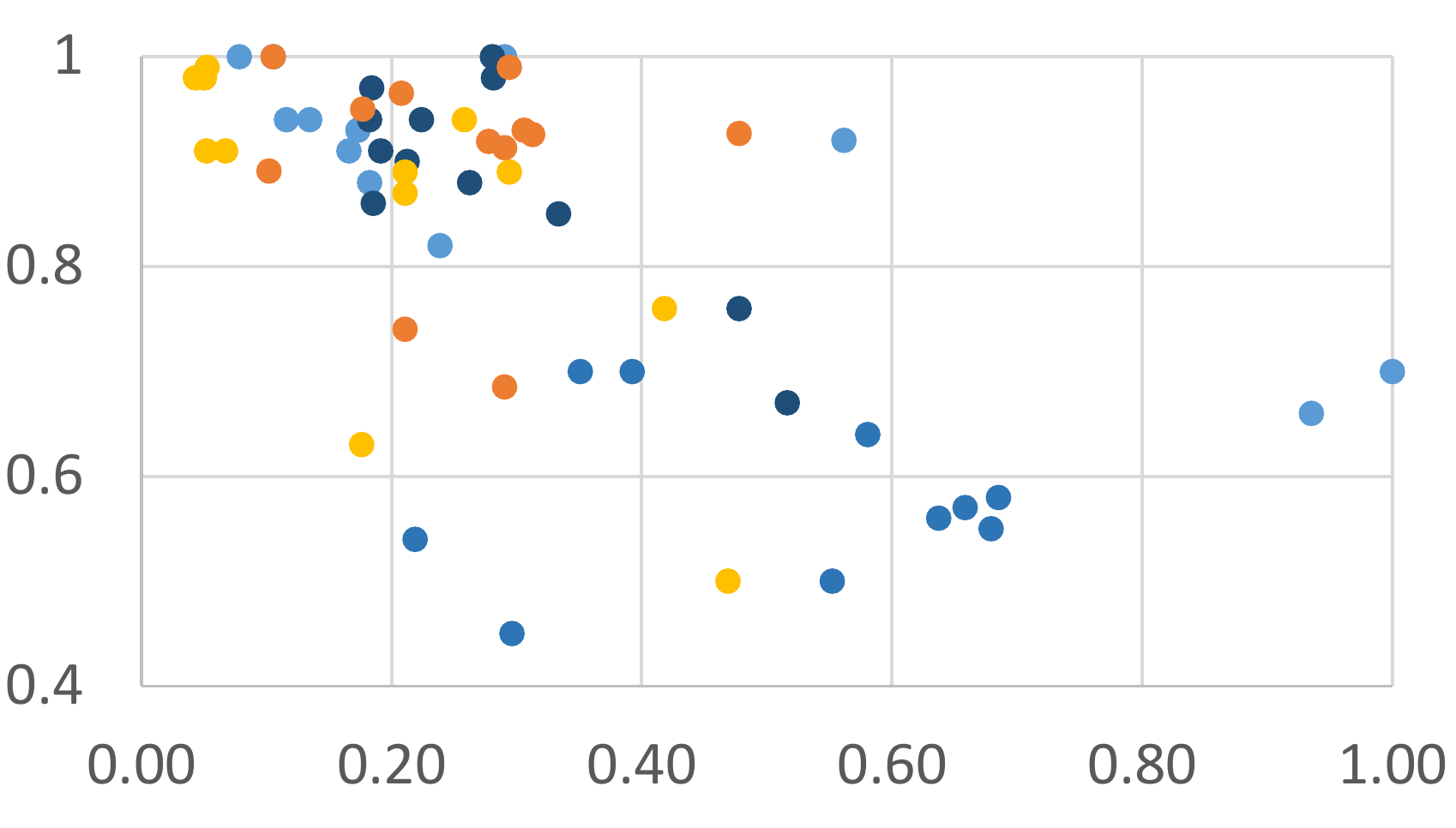}
    \includegraphics[height=2.9cm]{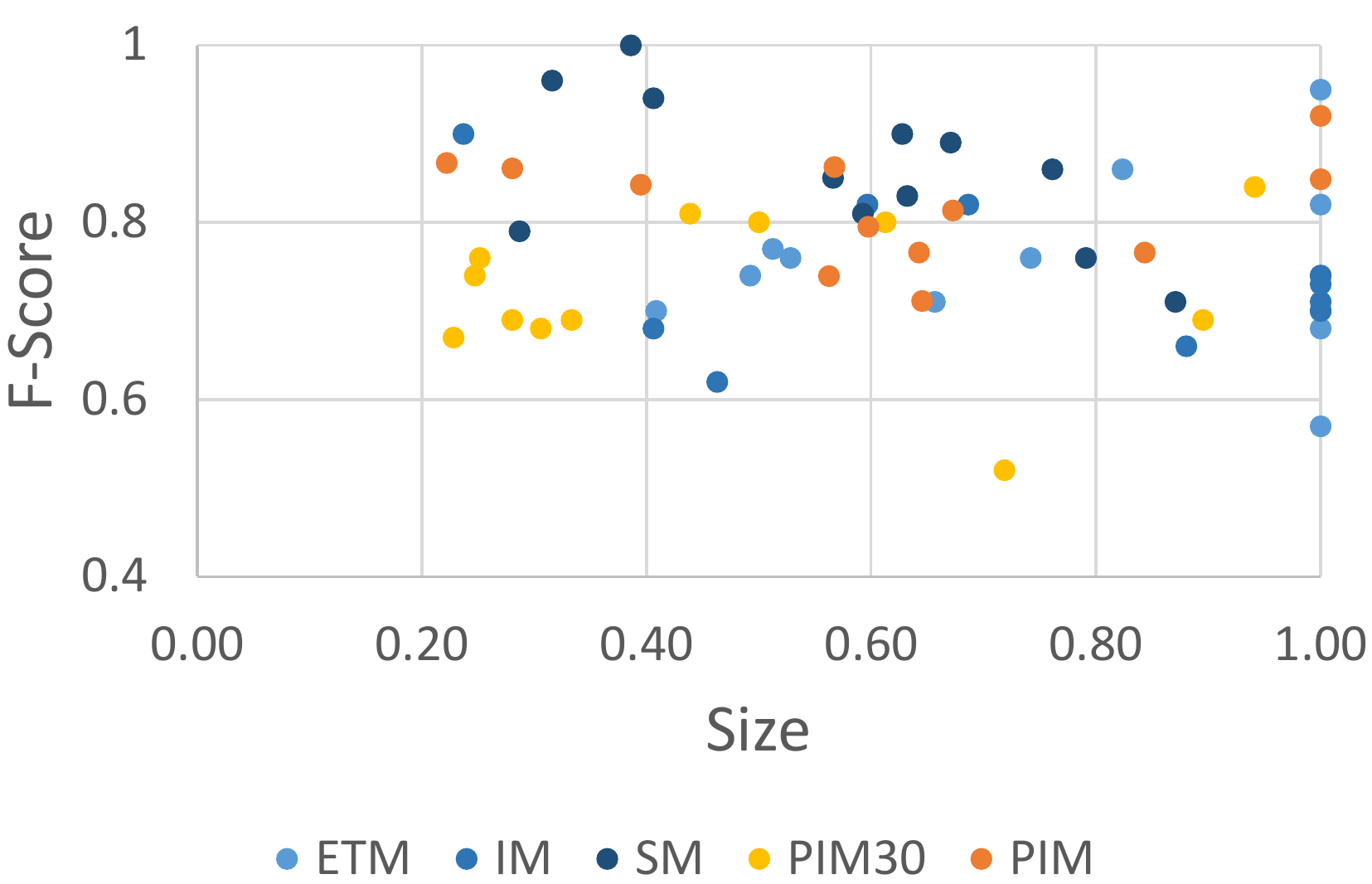}
    \includegraphics[height=2.9cm]{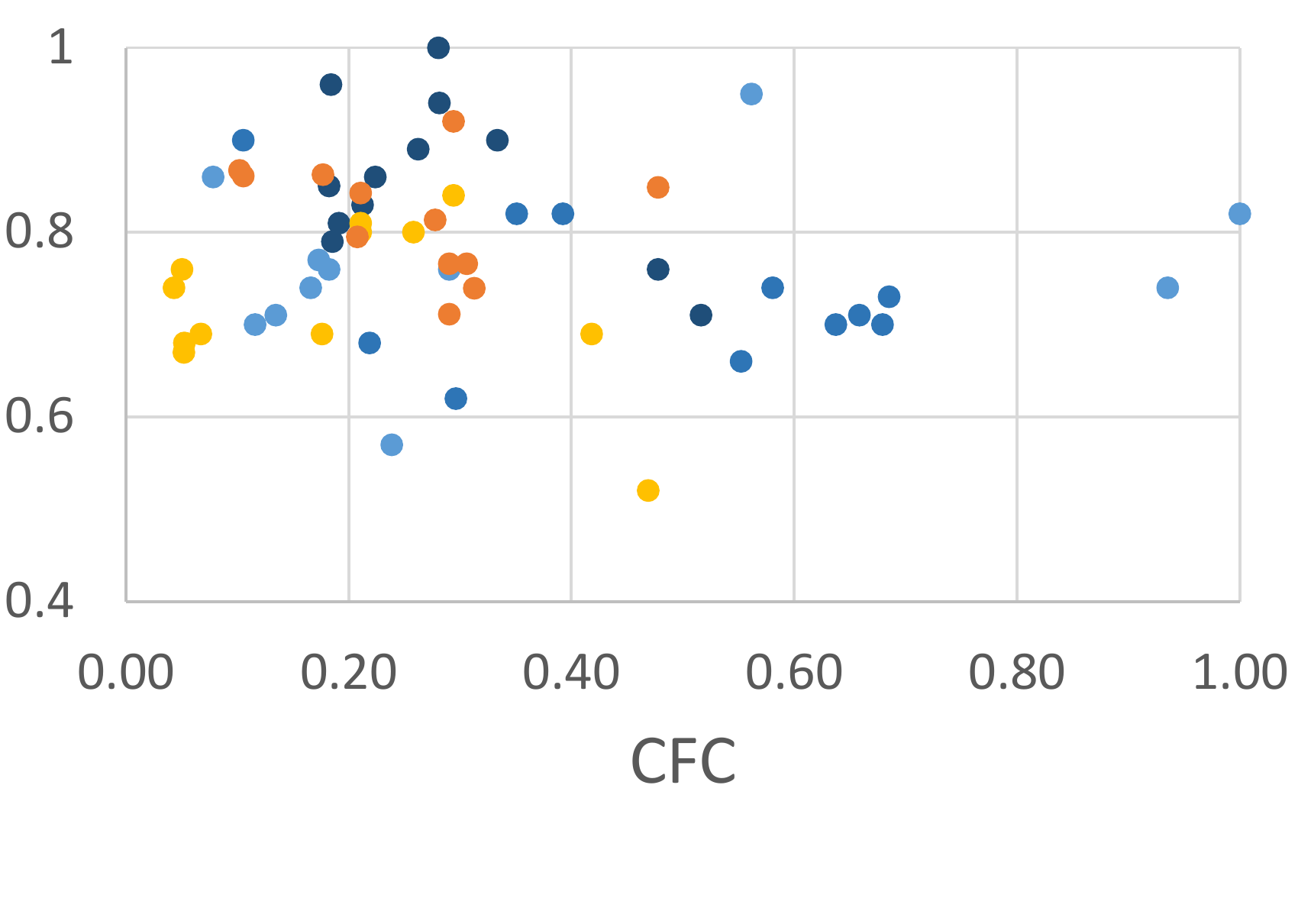}
    \caption{Fitness, precision, f-score vs normalized size and CFC of ETM, IMf, SM, PIM\textsubscript{30}, and PIM on benchmark.}
    \label{fig:eval:scatter1}
\end{figure}

\begin{figure}[t]
    \centering
    \includegraphics[height=2.9cm]{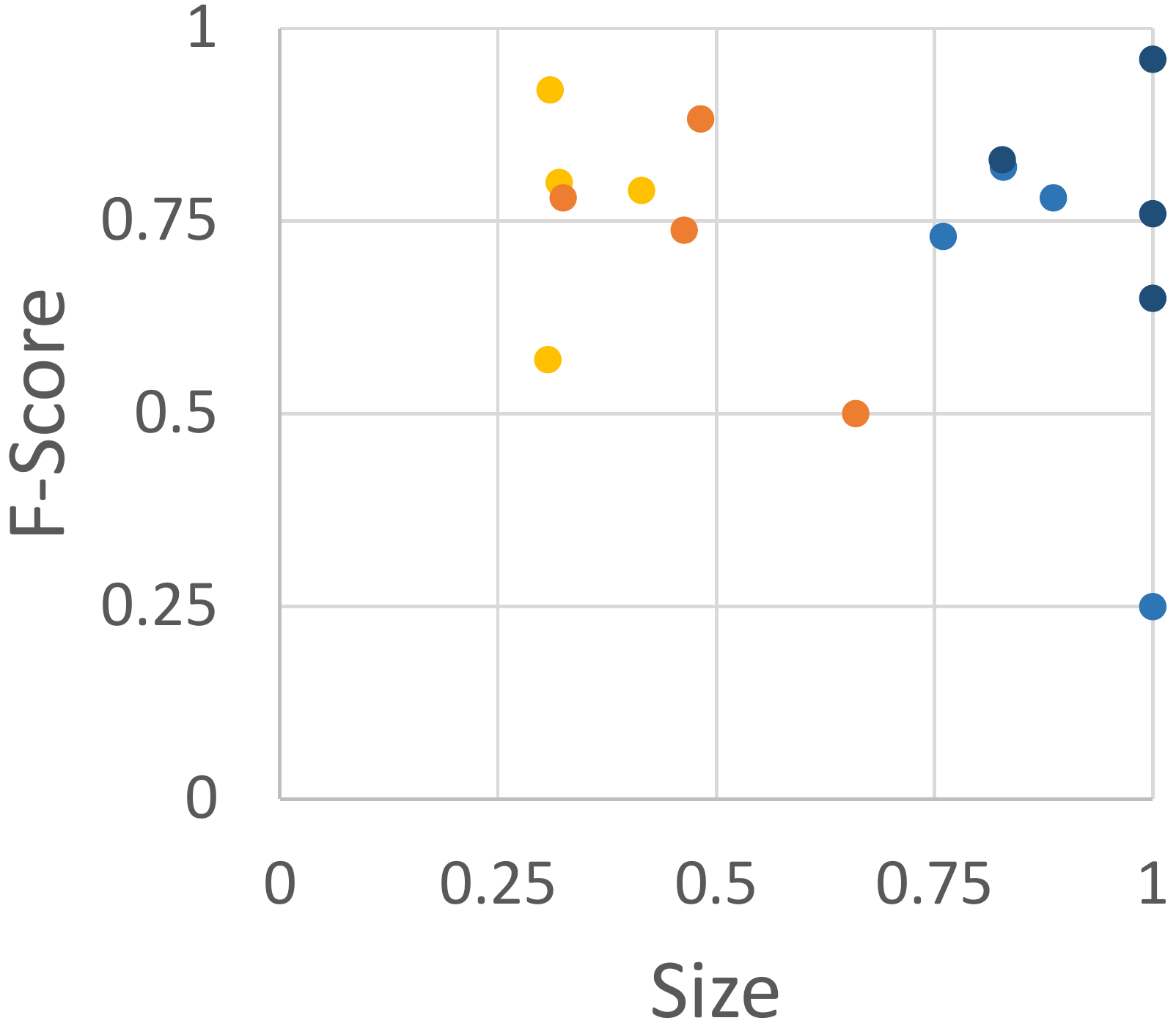}
    \includegraphics[height=2.9cm]{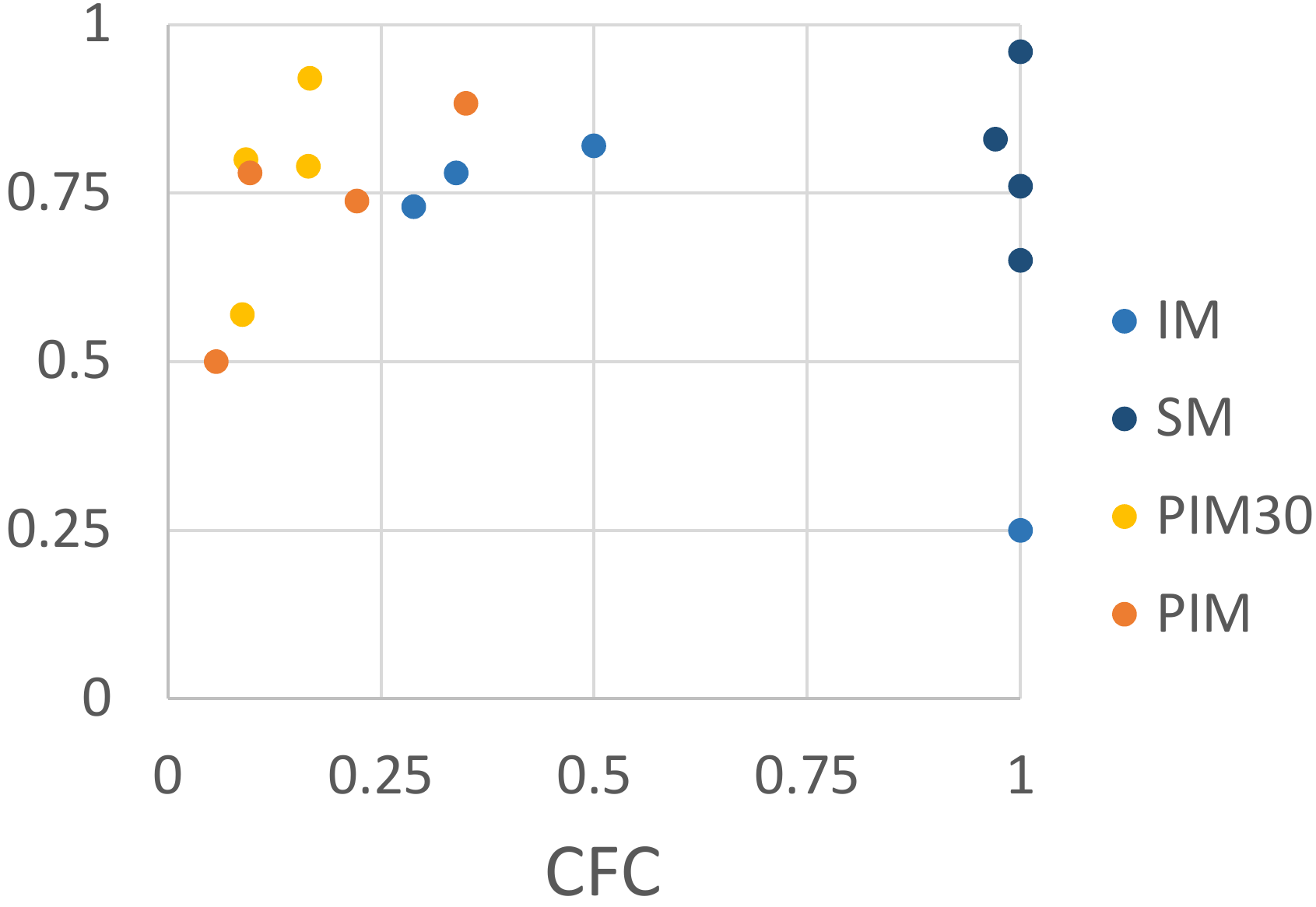}
    \caption{F-score vs normalized control-flow complexity of IMf, SM, PIM\textsubscript{30} on additional benchmark.}
    \label{fig:eval:scatter2}
\end{figure}

Figure~\ref{fig:results_all} shows fitness, precision, f-score, size, CFC, and running times for ETM, IMf, SM as reported in~\cite{Augusto2019AutomatedBenchmark} and for PIM\textsubscript{30} and PIM for all datasets; the Appendix shows individual measures. The scatter plots in Fig.~\ref{fig:eval:scatter1} and Fig.~\ref{fig:eval:scatter2} visualize fitness, precision, and F-score against size and CFC (normalized against the largest size/CFC measured per log).

On the benchmark event logs, PIM sacrifices fitness for increased precision (11/12 logs better than IM, 9/12 same or better than SM, 5/12 highest precision) and F-score (10/12 better than IMf, 4/12 highest, and 5/12 scoring 2nd best close to SM) while improving size (8/12 better than IMf, 4/12 better than SM) and CFC (10/12 better than IMf, 6/12 same or better than SM). PIM\textsubscript{30} sacrifices fitness even more and further reduces size and CFC due to activity filtering. PIM consistently shows higher fitness (12/12) and higher F-score (11/12) than PIM\textsubscript{30} but also shows higher precison on 7/12 logs. PIM's and PIM\textsubscript{30}'s results are more similar to those of ETM wrt. accuracy than to IMf and SM but finds larger models as well as smaller models than ETM does. The scatter plots in Fig.~\ref{fig:eval:scatter1} show that PIM strikes a novel balance in the pareto-front having more results with high precision and low size/CFC (top-left corner) compared to IMf and SM while the overall F-score for is comparable to all other techniques. PIM\textsubscript{30} either finds a solution within a second (5/12 logs) or takes several hours to complete whereas PIM completes in less than a second for all data sets (fastest for 11/12) due to the heuristics in \textsc{FindCut}.

On the 4 additional datasets, SM achieves highest fitness and PIM sacrifices fitness most. PIM\textsubscript{30} (and PIM) clearly outperforms IMf and SM in precision, size, and CFC on all (3/4) datasets, yielding F-scores comparable to IMf and SM (see Fig.~\ref{fig:results_all}); PIM models are slightly larger and less accurate than PIM\textsubscript{30} models. The scatter plots in Fig.~\ref{fig:eval:scatter2} reinforce the prior finding even stronger: PIM\textsubscript{30} and PIM reach a novel area in the pareto-front of quality criteria by striking a much better balance in precision and simplicity not reached by other techniques while retaining comparable F-scores.

\emph{Parameter sensitivity.}
To verify that this is not due to parameter choices, we evaluated how accuracy and size change when changing IMf's, SM's, and PIM's parameter for filtering infrequent behavior on the 4 additional datasets. Increasing filtering for SM traded fitness for precision which results in near-constant accuracy and in up to 50\% fewer edges for models returned by SM and corresponding reduction of CFC. However SM size and CFC remained significantly higher than that of PIM with $f=99.5\%$. Increasing filtering for IMf lead to unpredictable changes in accuracy and size; size and complexity always remained above PIM with $f=99.5\%$. Thus SM and IMf cannot reach the particular spot of simplicity vs accuracy reached by PIM. Lowering PIM's filtering parameter $f$ allowed to reduce model size and CFC down to the 2 most frequent activities; model size and complexity thereby falls off in a negative exponential curve as many infrequent edges in $\dfg$ and $\dfgSind$ that contribute to complexity appear to be ``equally infrequent'' and get filtered out together when lowering $f$. This confirms that PIM's models remain simpler also under filtering parameters. Moreover, both the effective filtering parameter and the option to consider only the $k$ most frequent activities in cut detection allow to control model detail and complexity, satisfying (R1).

\subsection{Empirical Evaluation}\label{section:evaluation:empirical}

\emph{Setup.} We performed an empirical evaluation to test whether the particular spot of high accuracy and simplicity taken by PIM indeed achieves R2: that the models produced by PIM only show structures for which there is significant evidence in the data. We provided participants (11 professionals, 4 researches, 2 students) with the following. (1) A visualization of the most frequent trace variants with activities color-coded as generated by Prom's ``Explore Event Log'' visualizer that could fit legibly on one A4 sheet for RTFMP (99\% of the log), Invoice (94\% of the log), and BPIC17\textsubscript{cp} (BPIC17\textsubscript{LC} filtered to only ``complete'' events, 55\% of the log). (2) Automatically laid-out diagrams of BPMN translations of the respective models produced by IMf, SM, and PIM (setup as in Sect.~\ref{section:evaluation:benchmark}), with algorithm anonymized and in random order. (3) Instructions to highlight with pens of two distinct colors all model structures they consider ``good'' (understand and have sufficient evidence in the data) or ``bad'' (do not understand or lack evidence). (4) A questionnaire to rank model preference and indicate their reasons. No time limit was given.

\begin{figure}[t]
    \centering
     \subfloat[RTFMP \label{subfig-choices:RTFMP}]{%
       \includegraphics[width=0.15\textwidth]{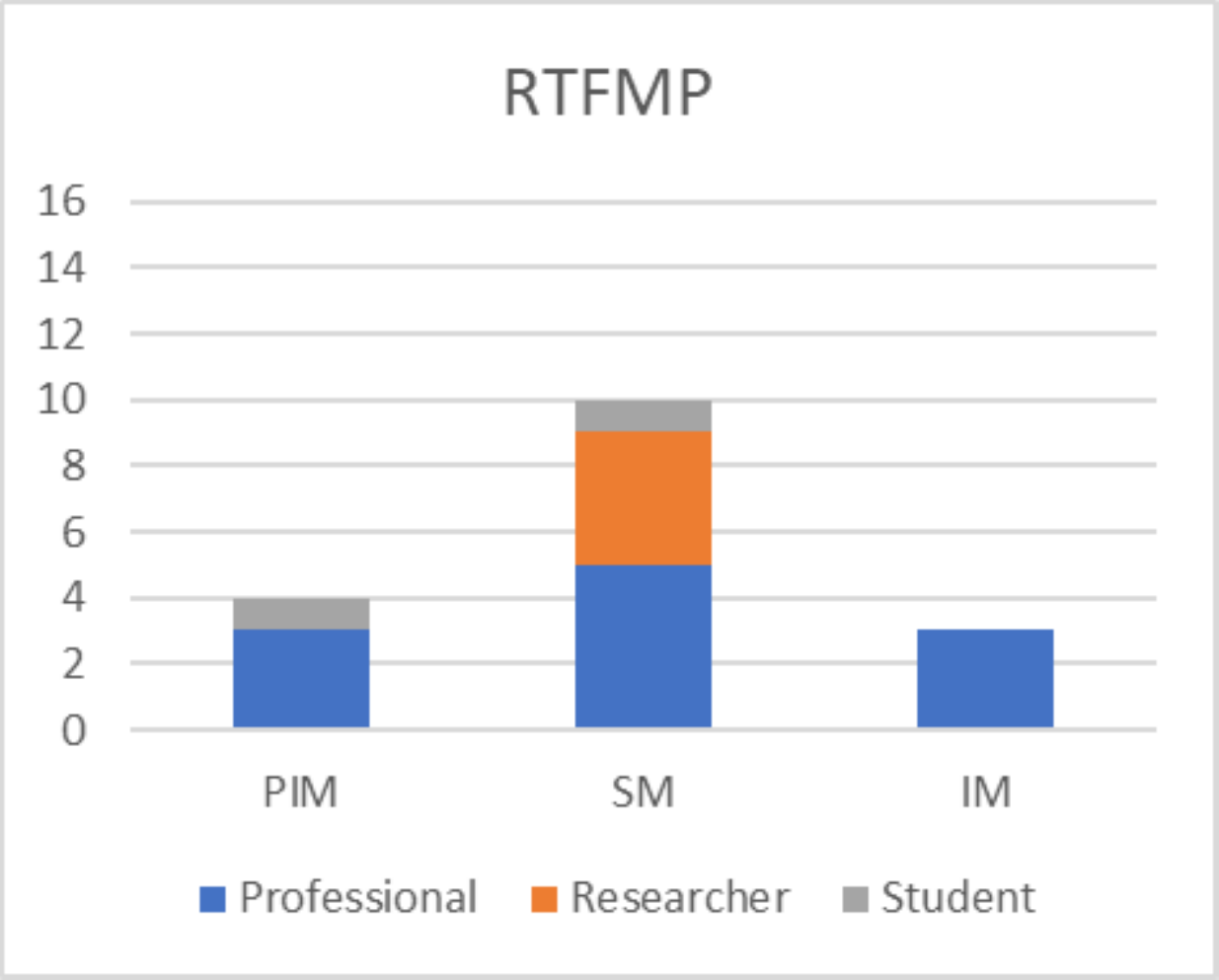}
     }
     \subfloat[Invoice \label{subfig-choices:Invoice}]{%
       \includegraphics[width=0.15\textwidth]{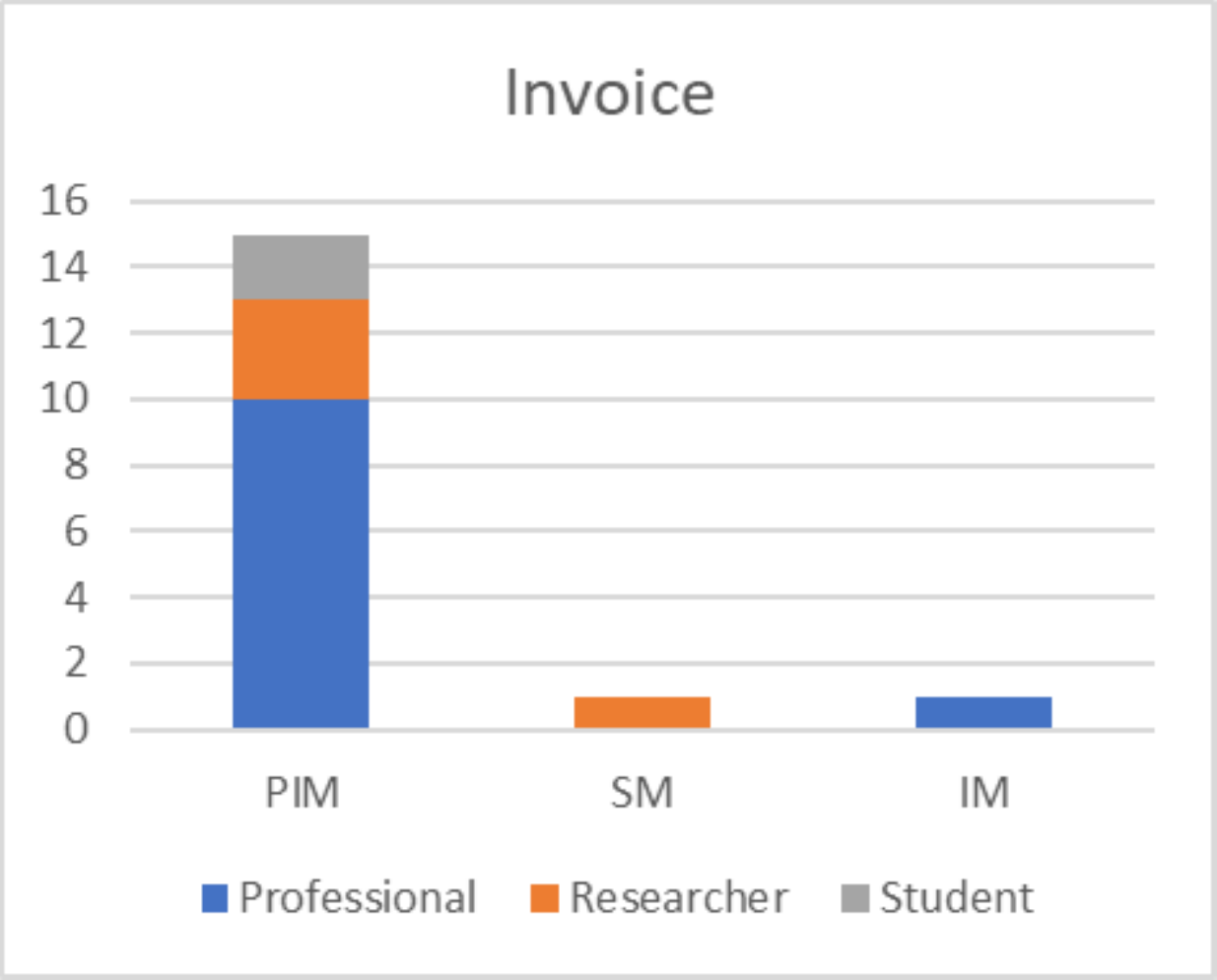}
     }
     \subfloat[BPIC17\textsubscript{cp}\label{subfig-choices:BPIC17}]{%
       \includegraphics[width=0.15\textwidth]{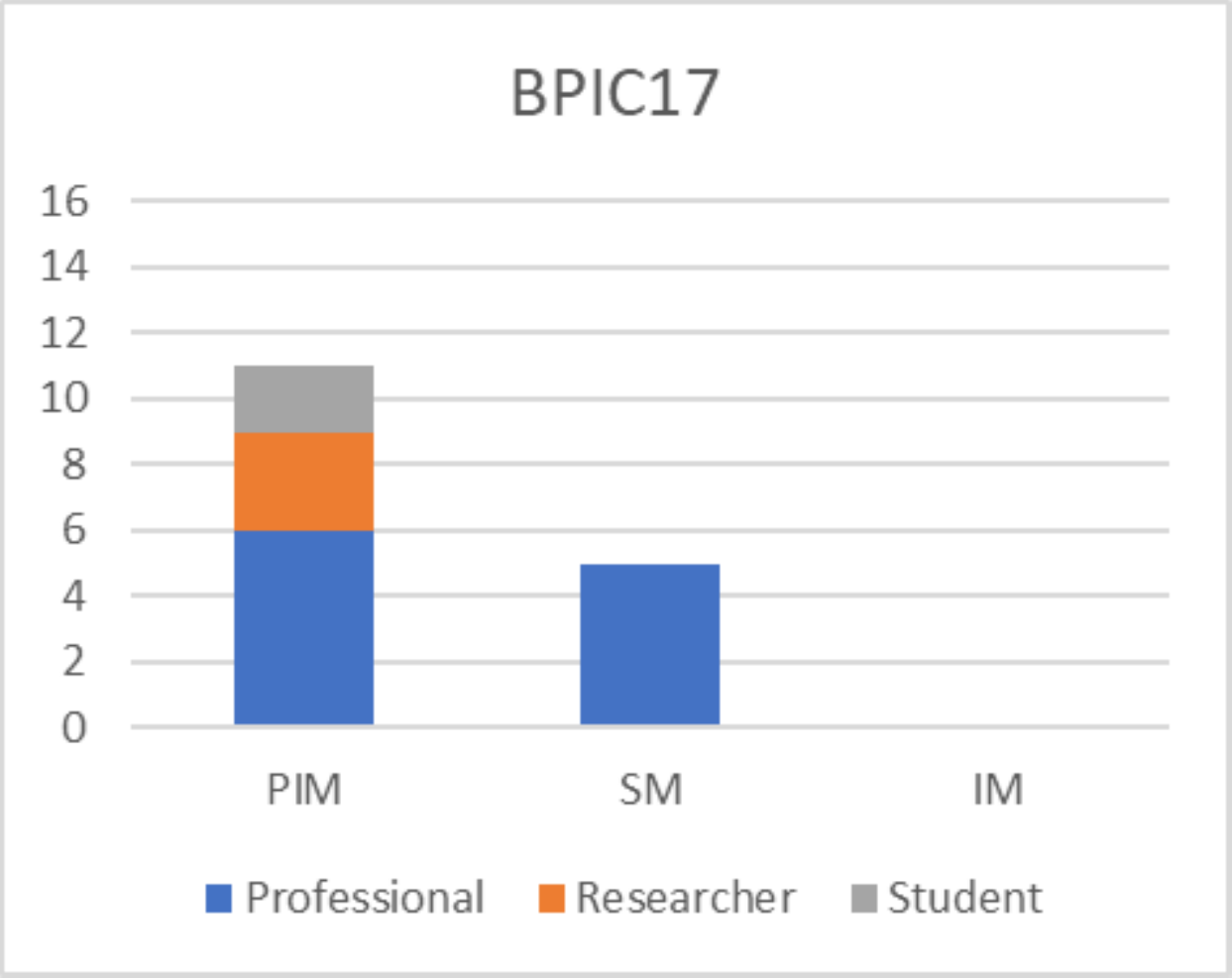}
     }

     \captionsetup{justification=centering}
     \caption{Favorite model choices made by the participants of the questionnaire}
     \label{eval:fig:choices}
   \end{figure}

\begin{figure}[t]
    \centering
    \subfloat[Good \label{subfig-good:BPIC17:PIM}]{%
       \includegraphics[width=0.108\textwidth]{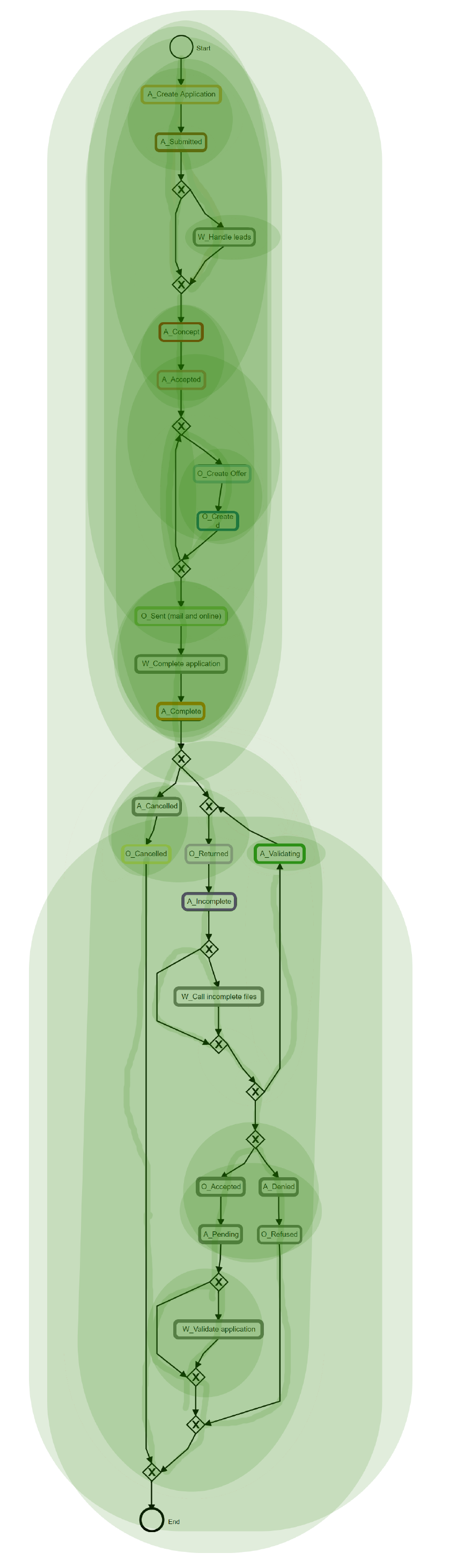}
    }
    \subfloat[Bad \label{subfig-bad:BPIC17:PIM}]{%
       \includegraphics[width=0.07\textwidth]{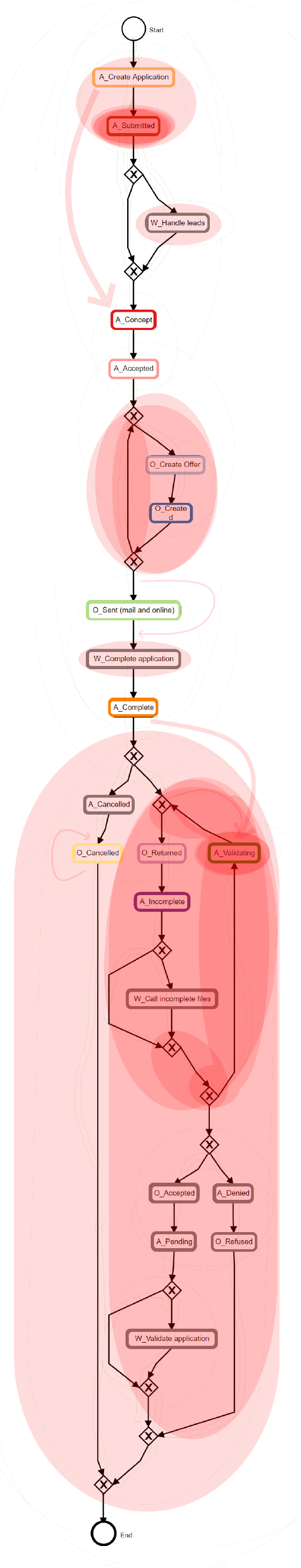}
    }
    \subfloat[Good \label{subfig-good:BPIC17:SM}]{%
       \includegraphics[width=0.15\textwidth]{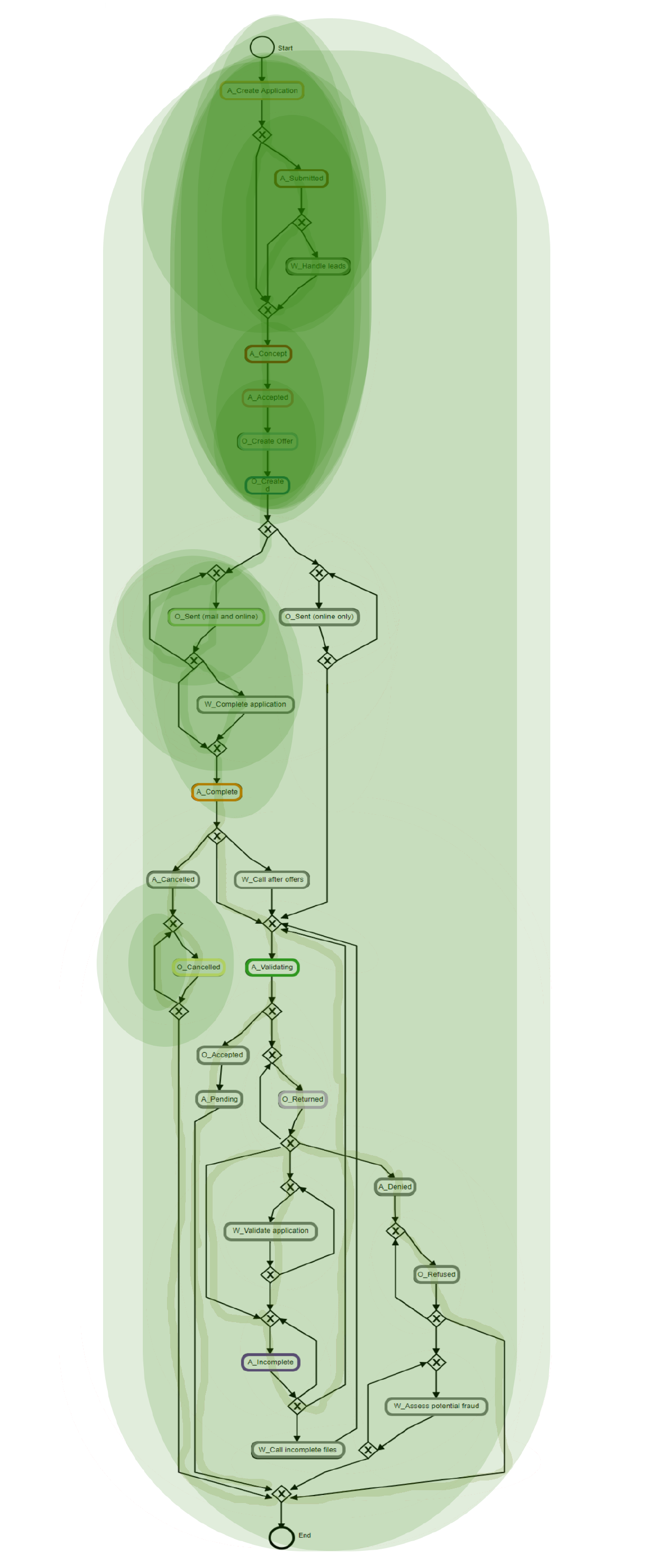}
    }
    \subfloat[Bad \label{subfig-bad:BPIC17:SM}]{%
       \includegraphics[width=0.13\textwidth]{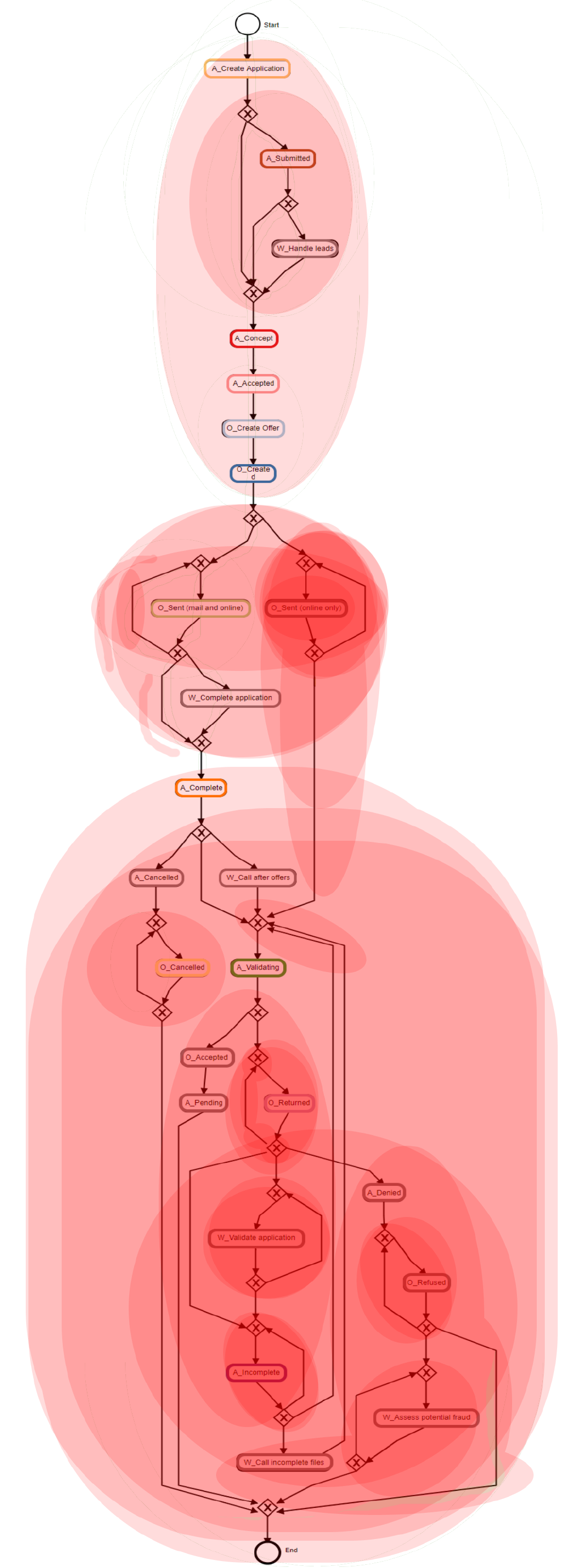}
    }

     \captionsetup{justification=centering}
     \caption{Markings made on the BPIC17\textsubscript{cp} models discovered by PIM (a, b) and SM (c, d).}
     \label{eval:fig:goodbad:BPIC17}
\end{figure}

\emph{Results.} Fig.~\ref{eval:fig:choices} shows how often the model of the respective algorithm was chosen as best model per dataset. Where SM was the most-preferred model for RTFMP, PIM was preferred most for Invoice and BPIC17\textsubscript{cp}. We aggregated the individual participants' highlighting of ``good'' and ``bad'' model structures into heat maps. Fig.~\ref{eval:fig:goodbad:BPIC17} shows the heat maps for the PIM and SM models of BPIC17\textsubscript{LC}; App.~\ref{sec-app:evaluation_detailed} shows all heat maps.

From the heat maps and questionnaires, we observed that the participants generally trusted and understood model structures produced by PIM more than model structures produced by SM and IMf (c.f. Fig 10b vs 10d). Models by PIM were sporadically described as representing too little of the data, but never too much of the data. Of the 51 written motivations justifying the participants' model ranking, 30 (58.8\%) justified their choice for PIM due to simplicity, clarity, or readability of the models. These results together confirm the particular usefulness of PIM's unique balance in fitness vs precision vs simplicity seen in Sect.~\ref{section:evaluation:benchmark}, satisfying R2.

\section{Conclusion} \label{section:conclusion}
We combined principles of the IMc~\cite{Leemans2014DiscoveringLogs} algorithm of the IM framework~\cite{Leemans2013DiscoveringApproach} with ideas from heuristic mining~\cite{Weijters2006ProcessAlgorithm} to design process discovery algorithm PIM that returns block-structured models of high simplicity and accuracy even on large, unfiltered event logs. PIM strikes a novel balance between higher precision, significantly lower model complexity, and comparable overall accuracy (F-score) compared to existing techniques; thereby reaching a new area in the pareto-front of model quality metrics. Our empirical evaluation confirms that this pareto-front area represents models that
are considered user-friendly and accurate for the data which was a central requirement raised by industrial analysts in a Delphi study prior to algorithm development. The models found by PIM are preferred over models produced by other state-of-the-art techniques; deviation analysis is then possible through model enhancement~\cite{Leemans2015ExploringDeviations} such as visual alignments~\cite{Bie2019VisualThesis} (c.f. App~\ref{sec-app:visual-alignment}).

\emph{Threats to validity.} The standard benchmark~\cite{Augusto2019AutomatedBenchmark} only contains filtered variants of real-life event logs changing the nature of the problem, which we offset by additional but fewer unfiltered event logs. All participants of the empirical evaluation were based in the Eindhoven area in Netherlands possibly introducing bias in preferences.

\emph{Future work.} Algorithmically, PIM seems to inherently favor precision over fitness while IMf's and SM's notable quality is ensuring fitness and high fitness/precision balance, respectively. It is worth exploring whether adding further parameters to PIM allows to produce fully fitting and precise models, possibly at the expense of simplicity.

However, in the qualitative evaluation, participants expressed a clear preference for simplicity and accuracy over fitness. This warrants reconsideration of the kinds of models automated process discovery techniques should produce. Process mining may need a focus shift: discovery of fully fitting models is no longer desirable if users do not find them usable.

\bibliographystyle{IEEEtran}
%

\bibliography{references}

\begin{appendices}
\section{Detailed Evaluation Results}\label{sec-app:evaluation_detailed}

\begin{table}[h!]

\footnotesize
\begin{tabular}{ | @{}C{3.3em} || @{}C{2.1em} | @{}C{1.6em} | @{}C{1.6em} | @{}C{1.6em} | @{}C{1.6em} | @{}C{1.6em} | @{}C{1.6em} | @{}C{1.2em} | @{}C{1.2em} | @{}C{4.5em} | }
\rot{75}{\textit{Log}} & \rot{75}{\textit{Method}} & \rot{75}{\textit{Fitness}} & \rot{75}{\textit{Precision}} & \rot{75}{\textit{F-score}} &  \rot{75}{\textit{Gen. 3-fold}} & \rot{75}{\textit{Size}} & \rot{75}{\textit{CFC}} & \rot{75}{\textit{Structured}} &\rot{75}{\textit{Sound}} & \rot{75}{\textit{Time(s)}}\\
\hline
\hline
\multirow{4}{*}{BPIC12} & IMf & \textbf{.98} & .50 & .66 & \textbf{.98} & 59 & 37 & \textbf{yes} & yes & 6.6  \\
& ETM & .44 & .82 & .57 & t/o & 67 & \textbf{16} & \textbf{yes} & yes & 14,400  \\
& SM & .75 & .76 & .76 & .75 & \textbf{53} & 32 & no & yes & \textbf{.58}  \\
& PIM$_{30}$ & .63 & .76* & .69* & .64 & 60 & 28* & \textbf{yes} & yes & 9330\\
& PIM & .78 & \textbf{.93}* & .\textbf{.85}* &  & 67 & 32* & \textbf{yes} & yes & .63\\
\hline
\multirow{4}{*}{BPIC13\textsubscript{cp}} & IMf & .82 &\textbf{1.00} & .90 & .82 & \textbf{9} & \textbf{4} & \textbf{yes} & yes & .1  \\
& ETM & \textbf{1.00} & .70 & .82 & t/o & 38 & 38 & \textbf{yes} & yes &  14,400  \\
& SM & .94 & .97 & \textbf{.96} & \textbf{.94} & 12 & 7 & \textbf{yes} & yes & .03  \\
& PIM$_{30}$ & .72 & .89 & .80 & .71 & 19 & 8 & \textbf{yes} & yes & \textbf{.01} \\
& PIM & .98 & .74 & .84 &     & 15 & 8 & \textbf{yes} & yes & \textbf{.01} \\
\hline
\multirow{4}{*}{BPIC13\textsubscript{inc}} & IMf & .92 & .54 & .68 & \textbf{.92} & \textbf{13} & \textbf{7} & \textbf{yes} & yes & 1.0  \\
& ETM & \textbf{1.00} & .51 & .68 & t/o & 32 & 144 & \textbf{yes} & yes & 14,400  \\
& SM & .91 & \textbf{.98} & \textbf{.94} & .91 & \textbf{13} & 9 & \textbf{yes} & yes & .23  \\
& PIM$_{30}$ & .55 & .50 & .52 & .54 & 23 & 15 & \textbf{yes} & yes & .1 \\
& PIM & .62 & .93* & .74* &     & 18 & 10 & \textbf{yes} & yes & \textbf{.07} \\
\hline
\multirow{4}{*}{BPIC14\textsubscript{f}} & IMf & \textbf{.89} & .64 & .74 & \textbf{.89} & 31 & 18 & \textbf{yes} & yes & 3.4  \\
& ETM & .61 & \textbf{1.00} & .76 & t/o & 23 & 9 & \textbf{yes} & yes & 14,400  \\
& SM & .76 & .67 & .71 & .76 & 27 & 16 & no & yes & .59  \\
& PIM$_{30}$ & .69 & .94* & \textbf{.80}* & .69 & \textbf{19}* & \textbf{8}* & \textbf{yes} & yes & .43\\
& PIM & .74 & .69* & .71 &     & 20* & 9* & \textbf{yes} & yes & \textbf{.07} \\
\hline
\multirow{4}{*}{BPIC15\textsubscript{1f}} & IMf & \textbf{.97} & .57 & .71 & \textbf{.96} & 164 & 108 & \textbf{yes} & yes & .6  \\
& ETM & .56 & .94 & .70 & t/o & 67 & 19 & \textbf{yes} & yes & 14,400  \\
& SM & .90 & .88 & \textbf{.89} & .90 & 110 & 43 & no & yes & .48  \\
& PIM$_{30}$ & .55 & .91*  & .69 & .55 & \textbf{46}* & \textbf{11}* & \textbf{yes} & yes & 8246 \\
& PIM & .79 & \textbf{.95}* & .86* &    & 93* & 29* & \textbf{yes} & yes & \textbf{.15} \\
\hline
\multirow{4}{*}{BPIC15\textsubscript{2f}} & IMf & \textbf{.93} & .56 & .70 & \textbf{.94} & 193 & 123 & \textbf{yes} & yes & .7  \\
& ETM & .62 & \textbf{.91} & .74 & t/o & 95 & 32 & \textbf{yes} & yes & 14,400  \\
& SM & .77 & .90 & \textbf{.83} & .77 & 122 & 41 & no & yes & \textbf{.25}  \\
& PIM$_{30}$ & .53 & \textbf{.91}* & .67& .53 & \textbf{44}* & \textbf{10}* & \textbf{yes} & yes & 7425 \\
& PIM & .66 & \textbf{.91}* & .77* &    & 124* & 56* & \textbf{yes} & yes & \textbf{.23} \\
\hline
\multirow{4}{*}{BPIC15\textsubscript{3f}} & IMf & \textbf{.95} & .55 & .70 & \textbf{.95} & 159 & 108 & \textbf{yes} & yes & 1.3  \\
& ETM & .68 & .88 & .76 & t/o & 84 & 29 & \textbf{yes} & yes & 14,400  \\
& SM & .78 & .94 & \textbf{.85} & .78 & 90 & 29 & no & yes & .36  \\
& PIM$_{30}$ & .62 & \textbf{.98}* & .76* & .62 & \textbf{40}* & \textbf{8}* & \textbf{yes} & yes & 7072\\
& PIM & .68 & .97* & .80* &    & 95* & 33* & \textbf{yes} & yes & \textbf{.19} \\
\hline
\multirow{4}{*}{BPIC15\textsubscript{4f}} & IMf & \textbf{.96} & .58 & .73 & \textbf{.96} & 162 & 111 & \textbf{yes} & yes & .7  \\
& ETM & .65 & .93 & .77 & t/o & 83 & 28 & \textbf{yes} & yes & 14,400  \\
& SM & .73 & .91 & \textbf{.81} & .73 & 96 & 31 & no & yes & .25 \\
& PIM$_{30}$ & .59 & \textbf{.98}* & .74* & .60 & \textbf{40}* & \textbf{7}* & \textbf{yes} & yes & 7331\\
& PIM & .73 & .92* & \textbf{.81}* &    & 109* & 45* & \textbf{yes} & yes & \textbf{.16} \\
\hline
\multirow{4}{*}{BPIC15\textsubscript{5f}} & IMf & \textbf{.94} & .18 & .30 & \textbf{.94} & 134 & 95 & \textbf{yes} & yes & 1.5  \\
& ETM & .57 & .94 & .71 & t/o & 88 & 18 & \textbf{yes} & yes & 14,400  \\
& SM & .79 & .94 & \textbf{.86} & .79 & 102 & 30 & no & yes & .27  \\
& PIM$_{30}$ & .52 & \textbf{.99}* & .68* & .51 & \textbf{41}* & \textbf{7}* & \textbf{yes} & yes & 7061\\
& PIM & .65 & .93* & .77* &    & 113* & 41* & \textbf{yes} & yes & \textbf{.19} \\
\hline
\multirow{4}{*}{BPIC17\textsubscript{f}} & IMf & \textbf{.98} & .70 & .82 & \textbf{.98} & 35 & 20 & \textbf{yes} & yes & 13.3  \\
& ETM & .76 & \textbf{1.00} & .86 & t/o & 42 & \textbf{4} & \textbf{yes} & yes & 14,400  \\
& SM & .95 & .85 & .90 & .95 & \textbf{32} & 17 & no & yes & 2.53  \\
& PIM$_{30}$ & .80 & .89* & .84* & .80 & 48 & 15* & \textbf{yes} & yes & 8392 \\
& PIM & .86 & .99 & \textbf{.92}* &   & 51 & 15* & \textbf{yes} & yes & \textbf{1.01} \\
\hline
\multirow{4}{*}{RTFMP} & IMf & \textbf{.99} & .70 & .82 & .99 & 34 & 20 & \textbf{yes} & yes & 10.9  \\
& ETM & \textbf{.99} & .92 & .95 & t/o & 57 & 32 & \textbf{yes} & yes & 14,400 \\
& SM & \textbf{.99} & \textbf{1.00} & \textbf{1.00} & \textbf{1.00} & 22 & 16 & no & yes & 1.25  \\
& PIM$_{30}$ & .76 & .87* & .81& .76 & 25* & 12* & \textbf{yes} & yes & .011\\
& PIM & .76 & \textbf{1.00} & .86 &   & \textbf{16}* & \textbf{6}* & \textbf{yes} & yes & \textbf{.01}\\
\hline
\multirow{4}{*}{SEPSIS} & IMf & \textbf{.99} & .45 & .62 & \textbf{.96} & 50 & 32 & \textbf{yes} & yes & .4  \\
& ETM & .83 & .66 & .74 & t/o & 108 & 101 & \textbf{yes} & yes & 14,400  \\
& SM & .73 & .86 & .79 & .73 & 31 & 20 & no & yes & .05  \\
& PIM$_{30}$ & .77 & .63* & .69* & .77 & 36* & 19* & \textbf{yes} & yes & .463\\
& PIM & .84 & \textbf{0.89}* & \textbf{0.87}* &  & \textbf{24}* & \textbf{11}* & \textbf{yes} & yes & \textbf{.03}\\
\hline
\end{tabular}
\caption{Evaluation results on IMf, ETM, SM, PIM$_{30}$, and PIM; \textbf{bold} = best value per log; * = PIM improvement over IMf.}
\label{tab-app:results}
\end{table}

\begin{figure}[h!]
    \centering
    \subfloat[PIM\textsubscript{30}]{
        \includegraphics[width=.15\linewidth]{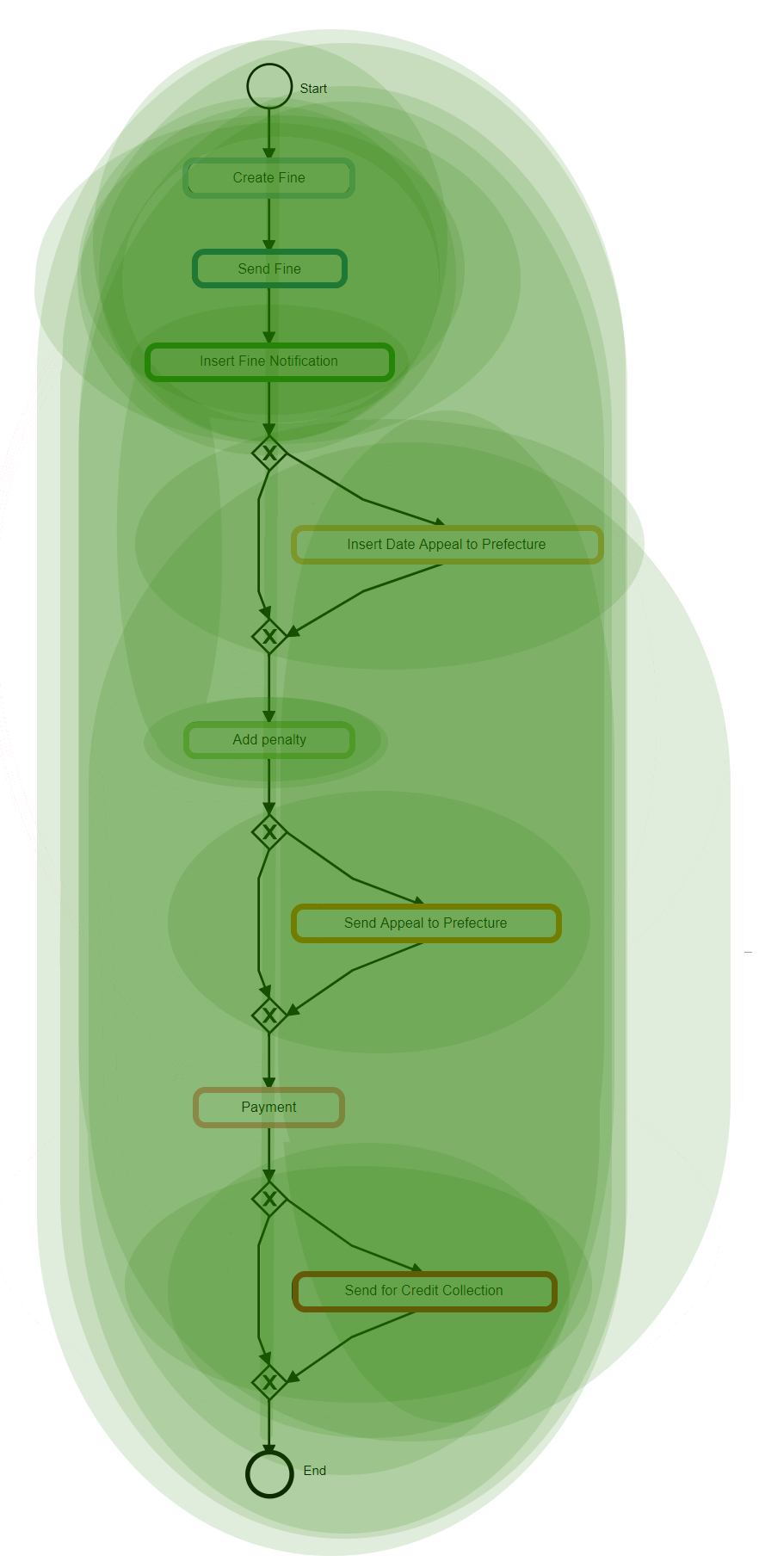}
        \includegraphics[width=.15\linewidth]{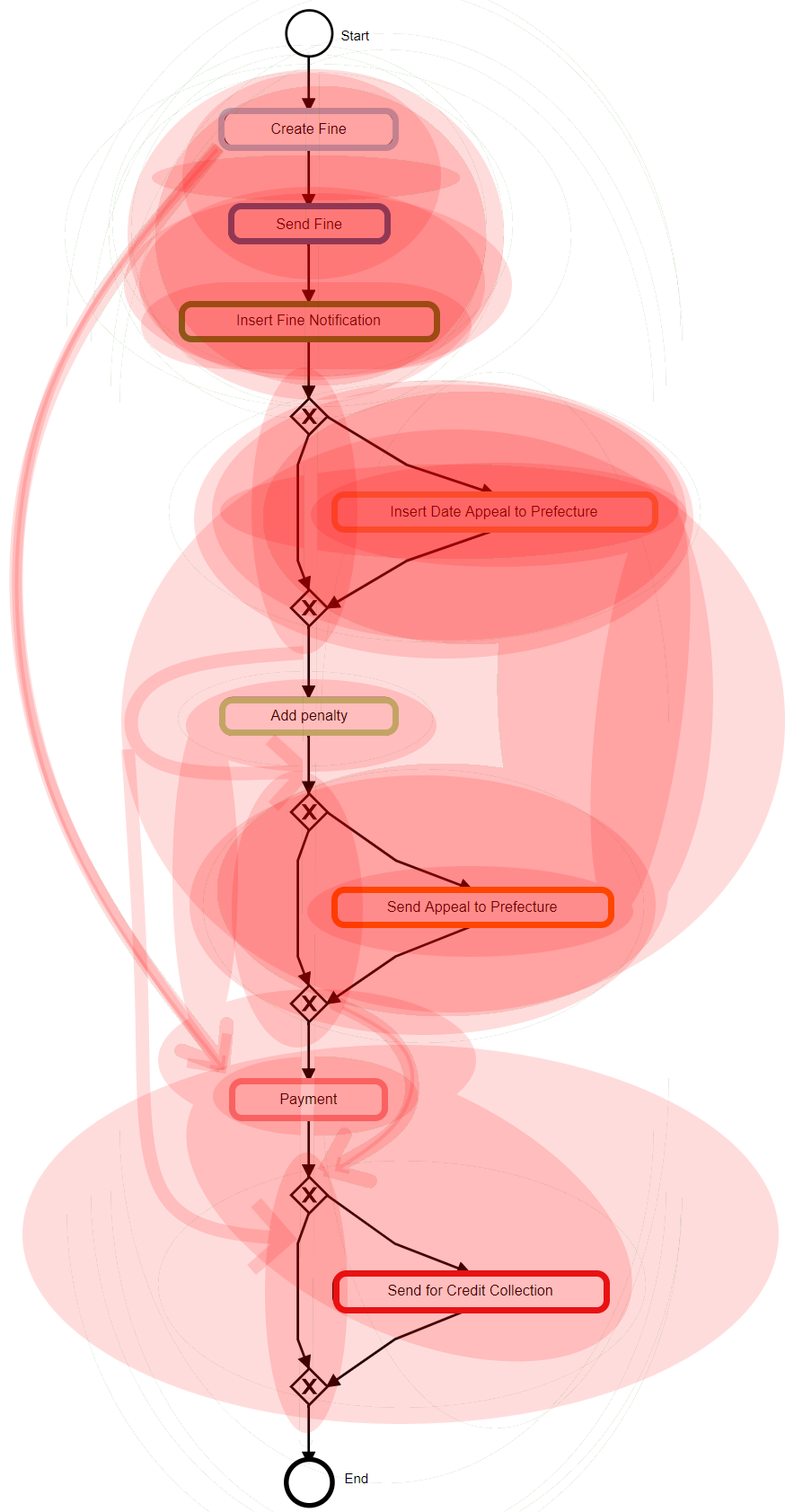}}
    \subfloat[IMf]{
        \includegraphics[width=.15\linewidth]{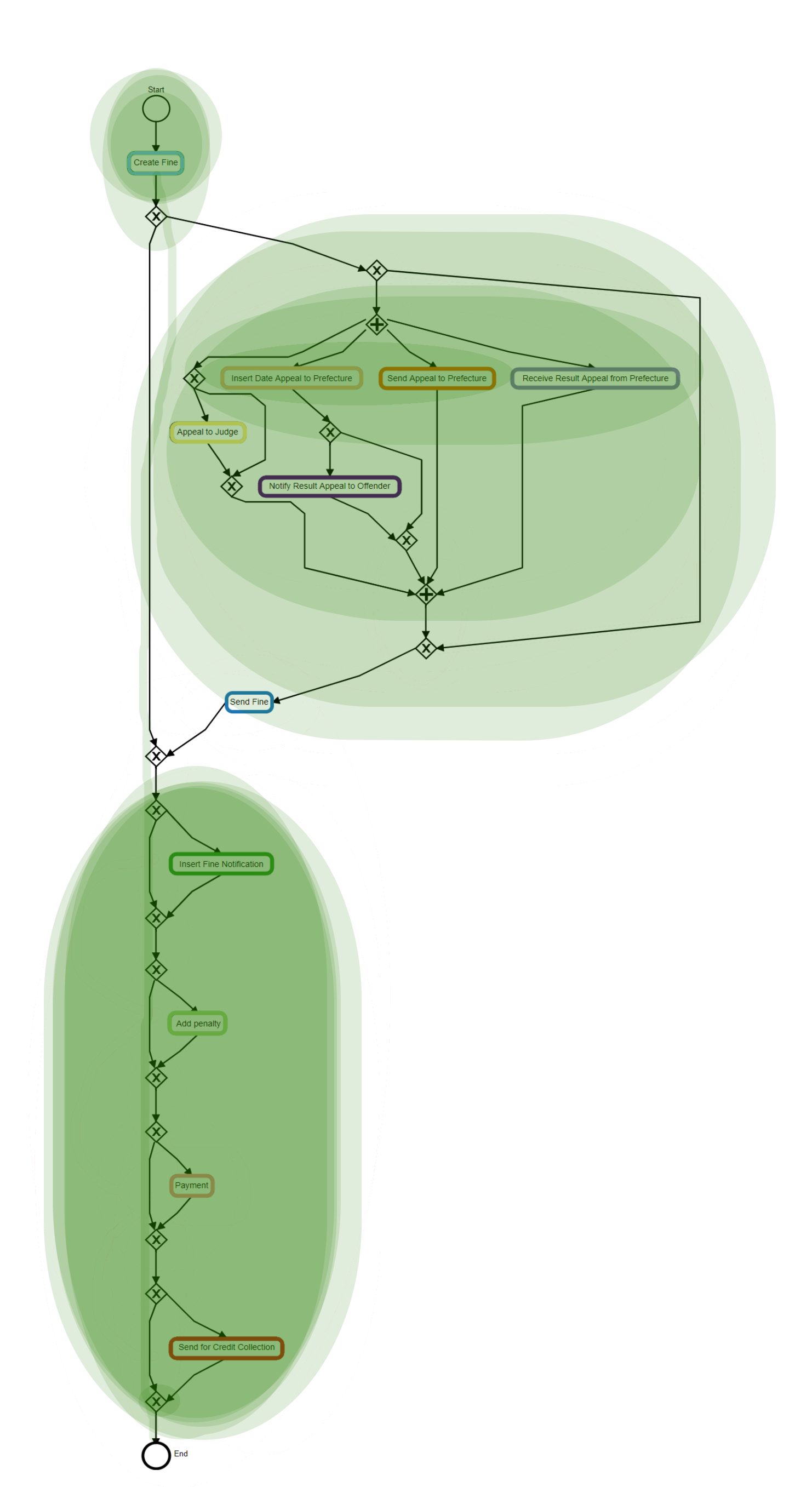}
        \includegraphics[width=.15\linewidth]{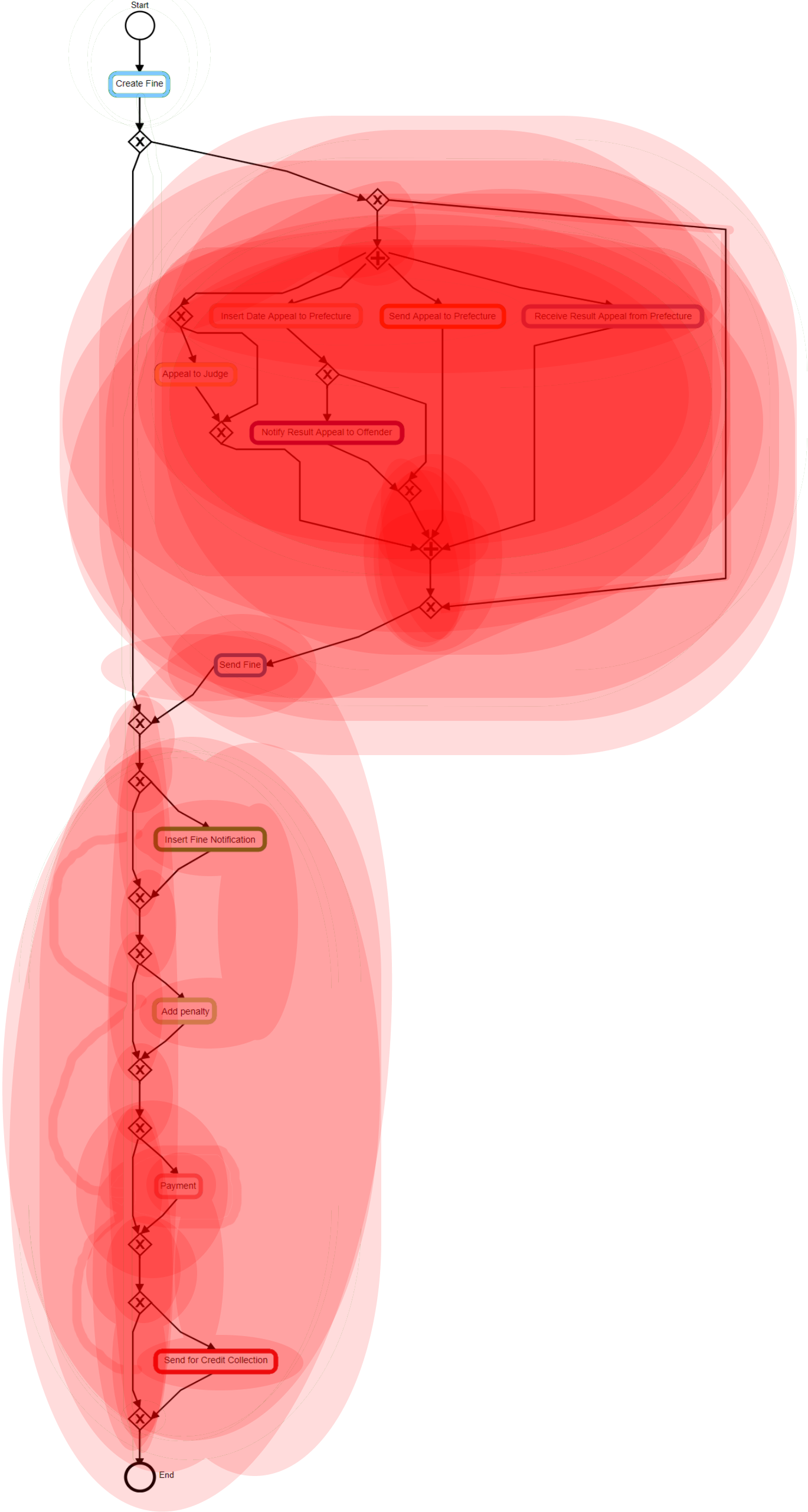}
    }
    \subfloat[SM]{
        \includegraphics[width=.15\linewidth]{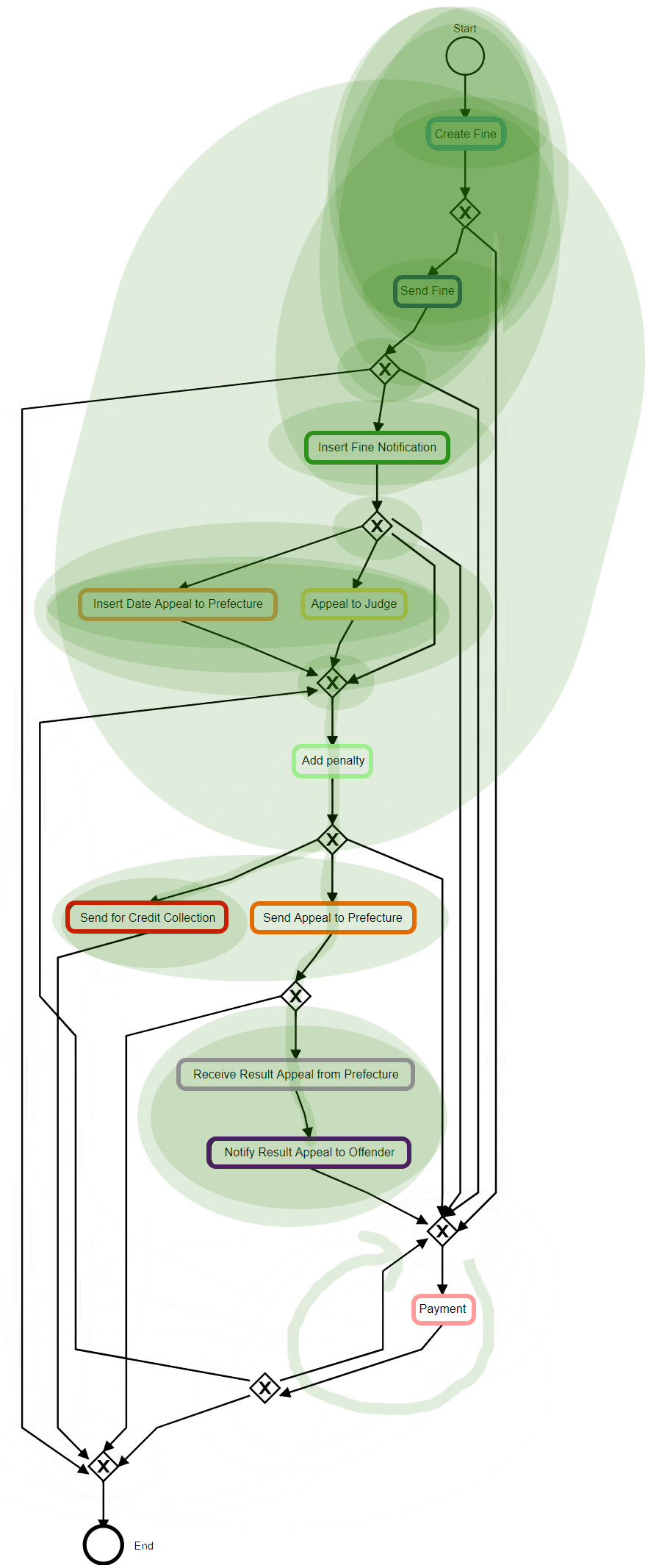}
        \includegraphics[width=.15\linewidth]{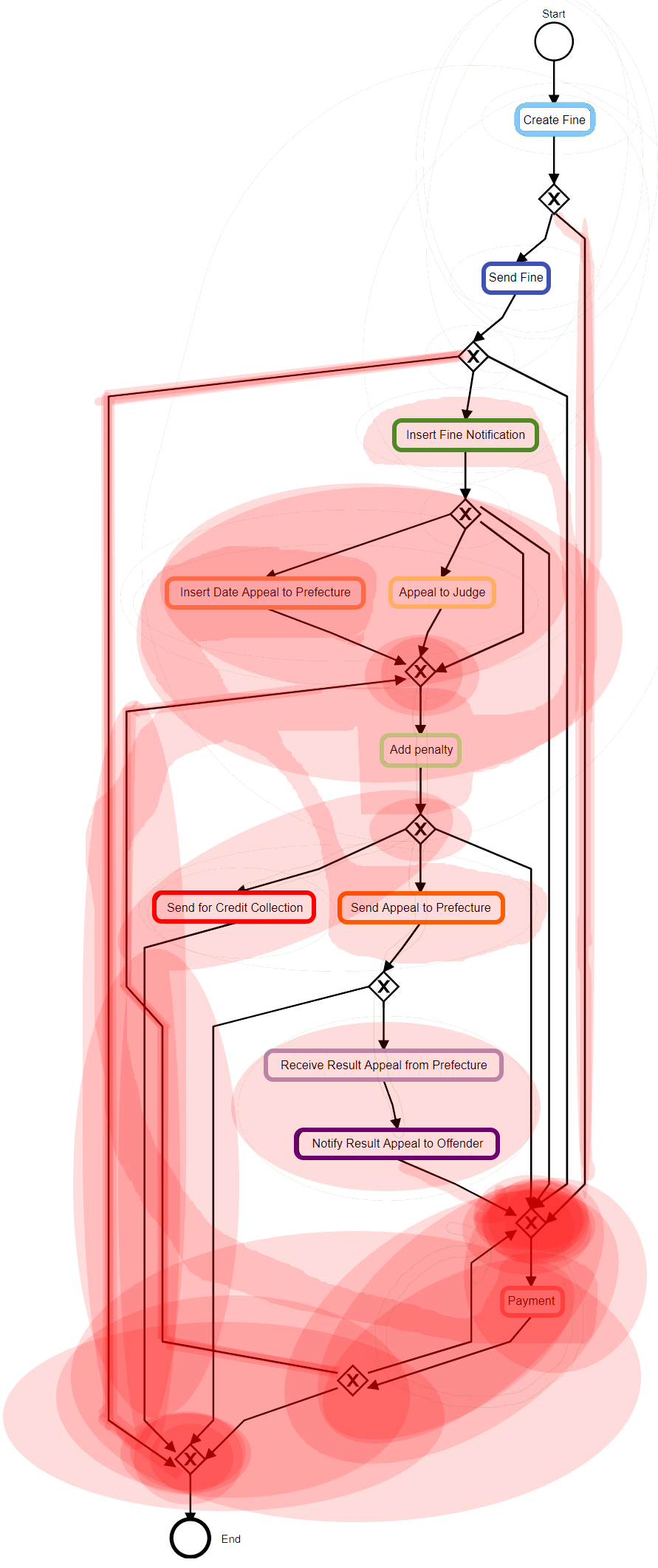}
    }
    \par
    \subfloat[PIM\textsubscript{30}]{
        \includegraphics[width=.15\linewidth]{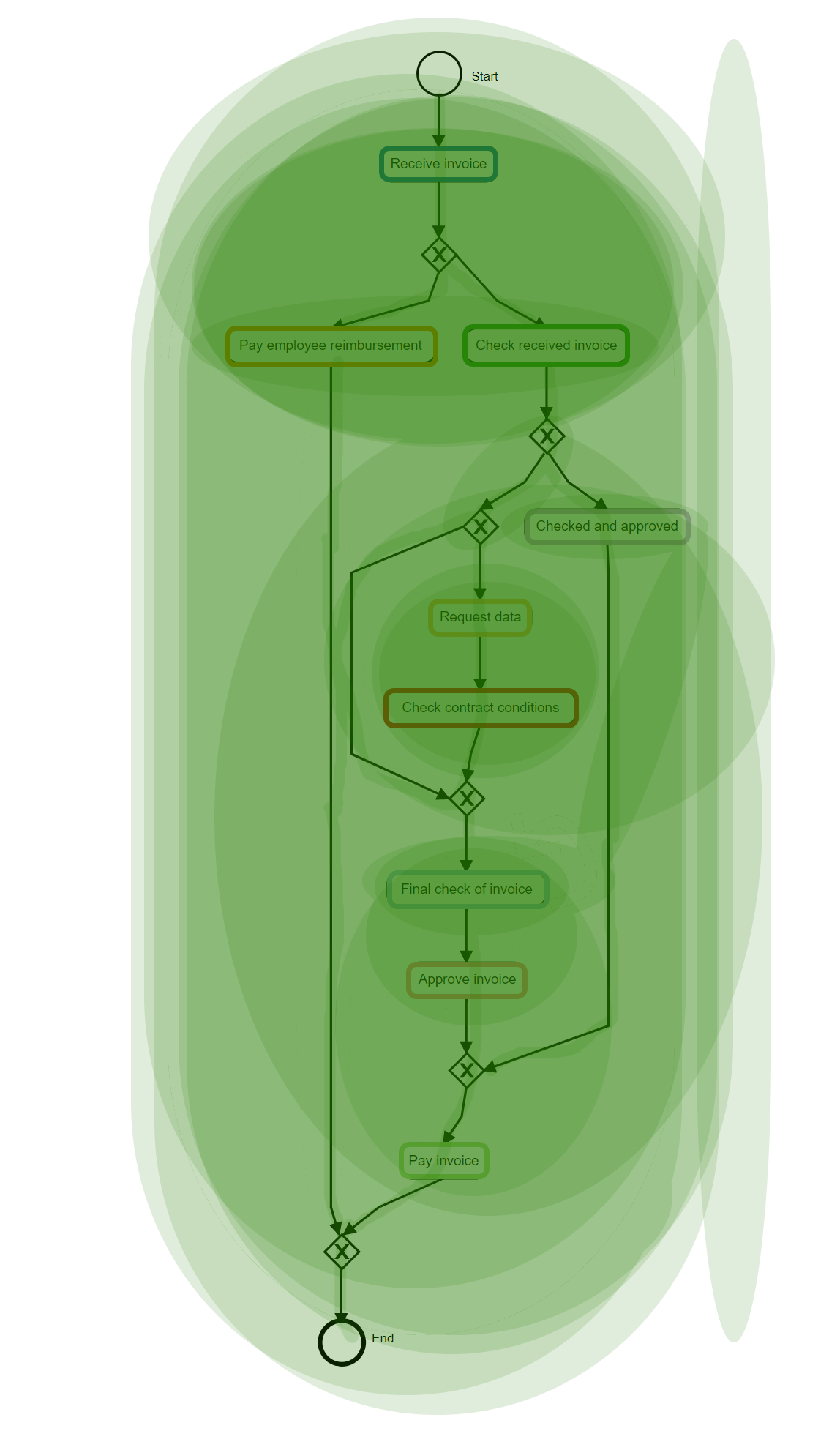}
        \includegraphics[width=.15\linewidth]{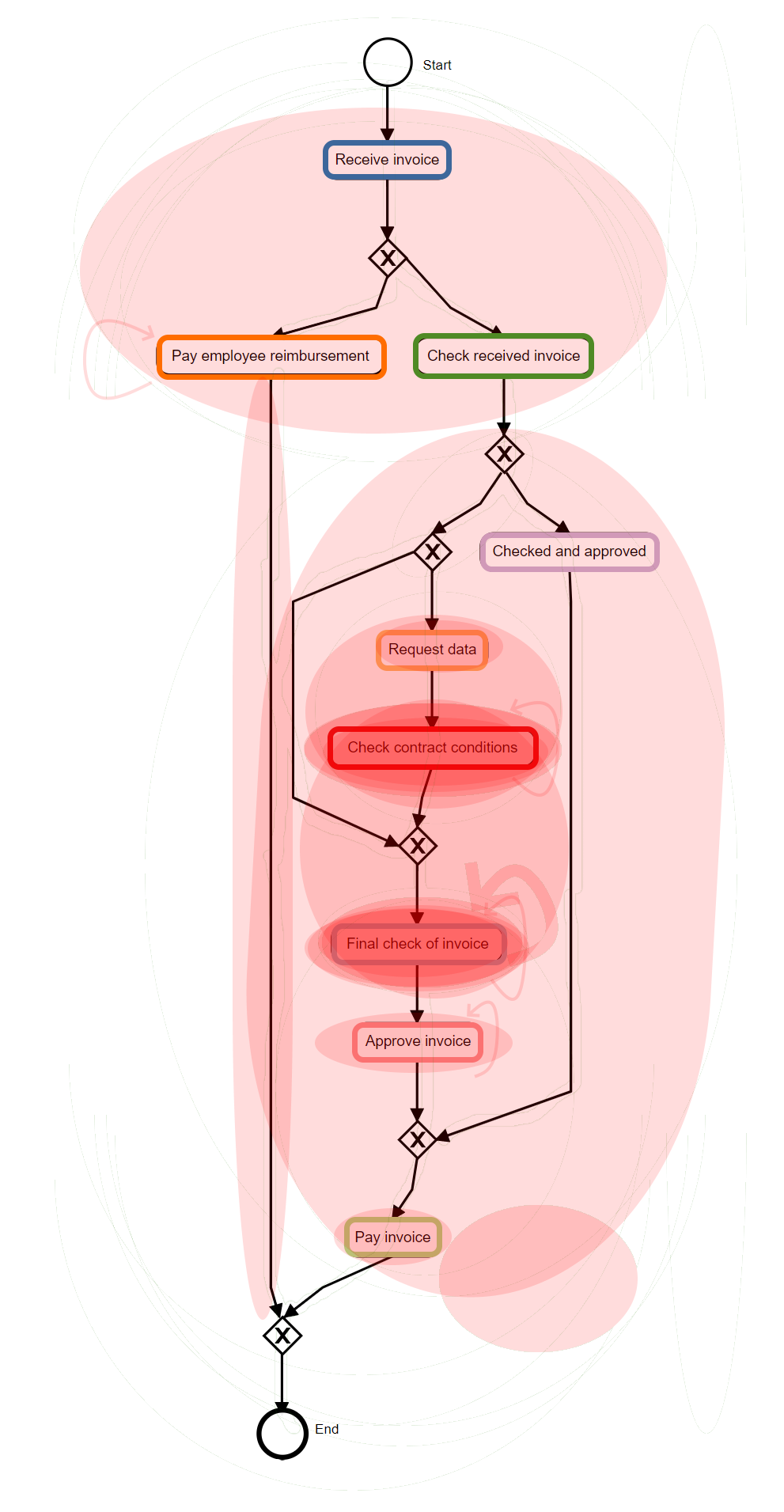}}
    \subfloat[IMf]{
        \includegraphics[width=.15\linewidth]{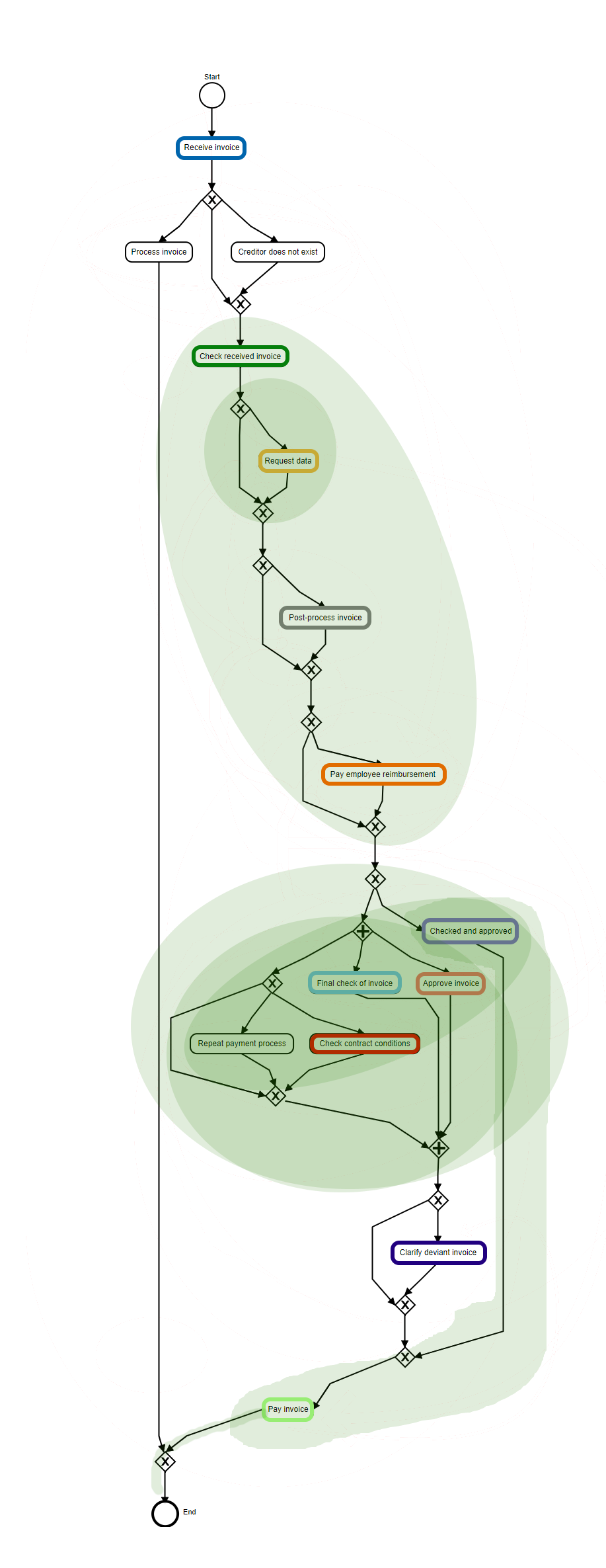}
        \includegraphics[width=.15\linewidth]{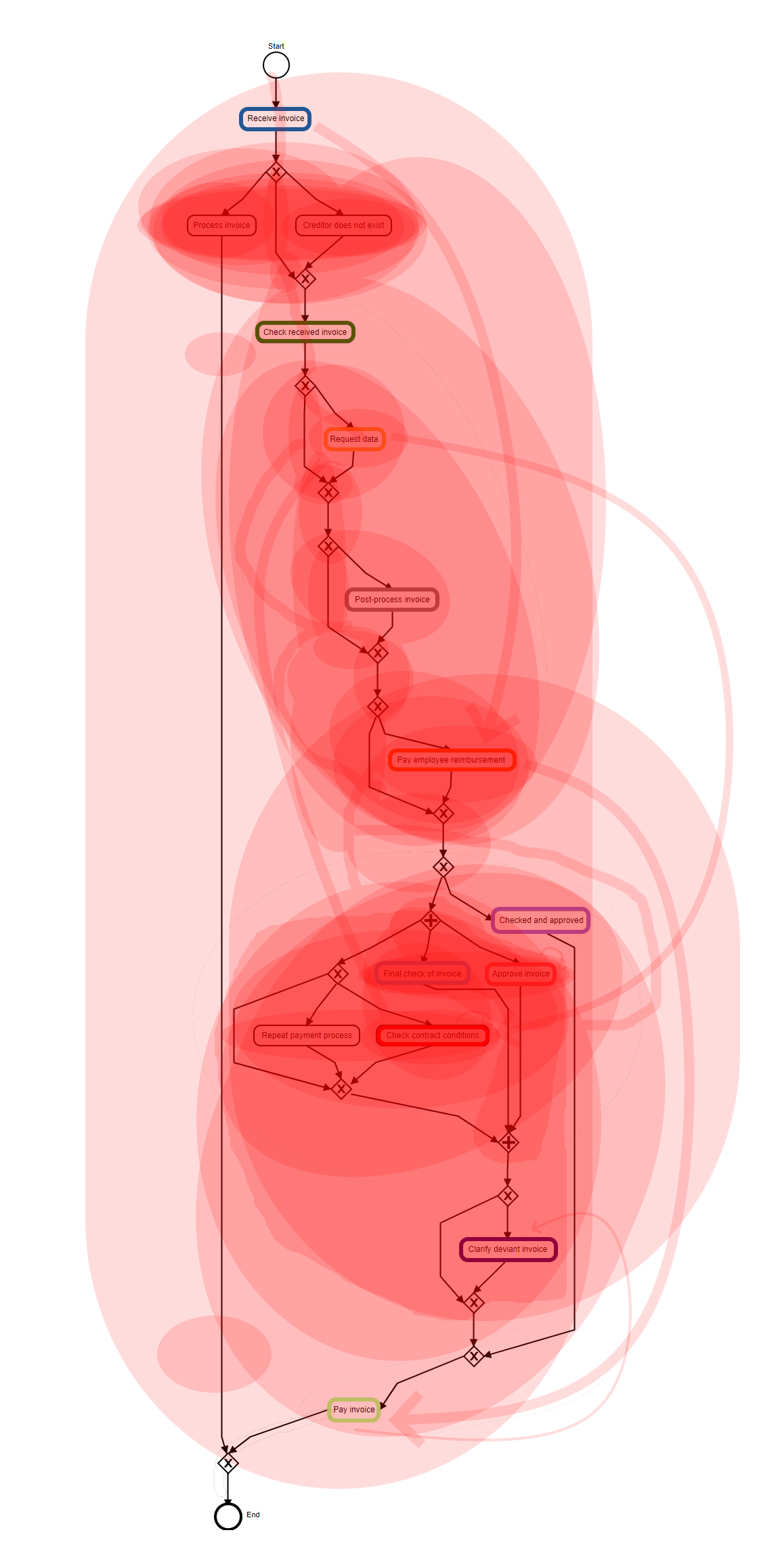}
    }
    \subfloat[SM]{
        \includegraphics[width=.15\linewidth]{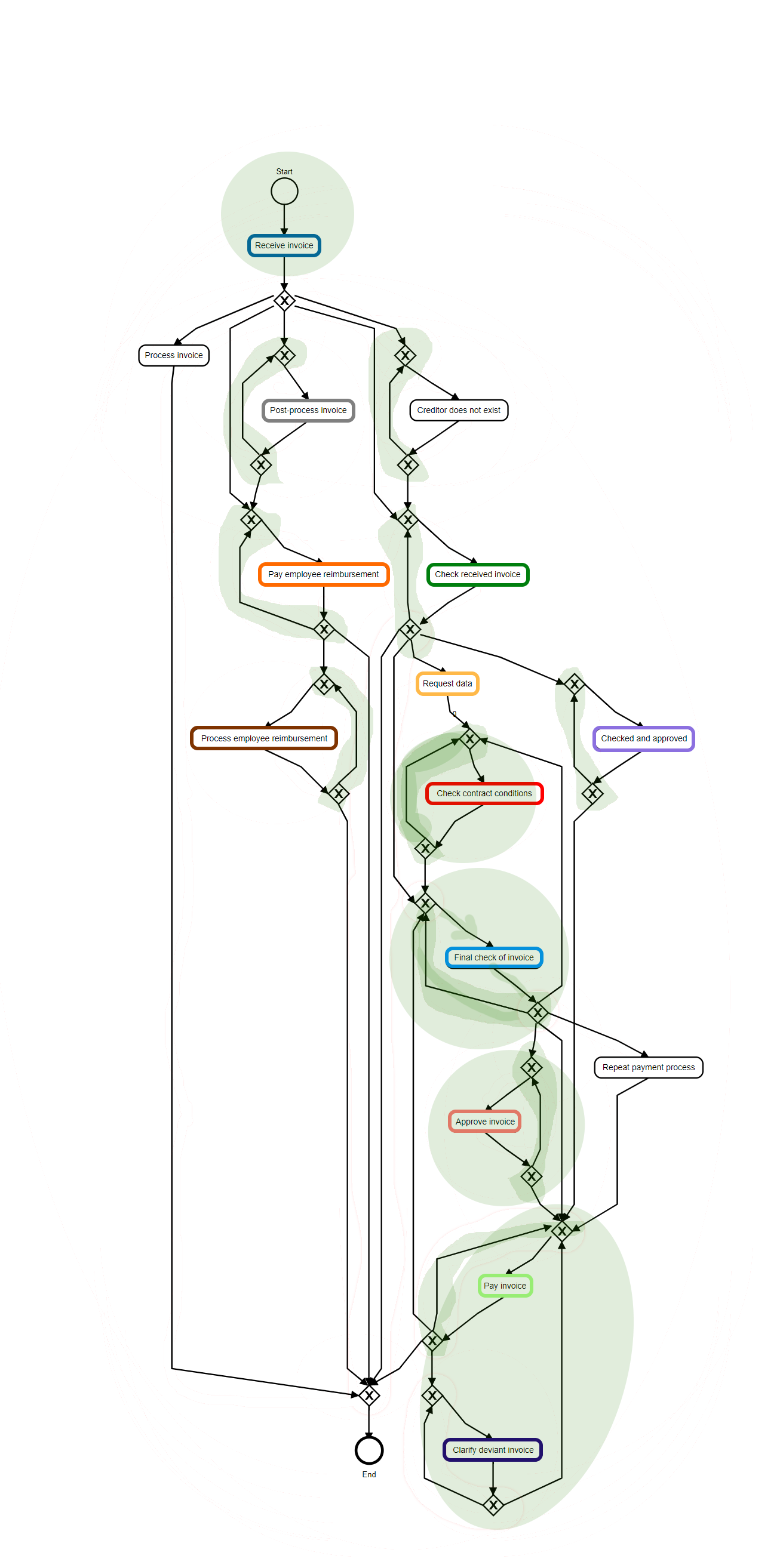}
        \includegraphics[width=.15\linewidth]{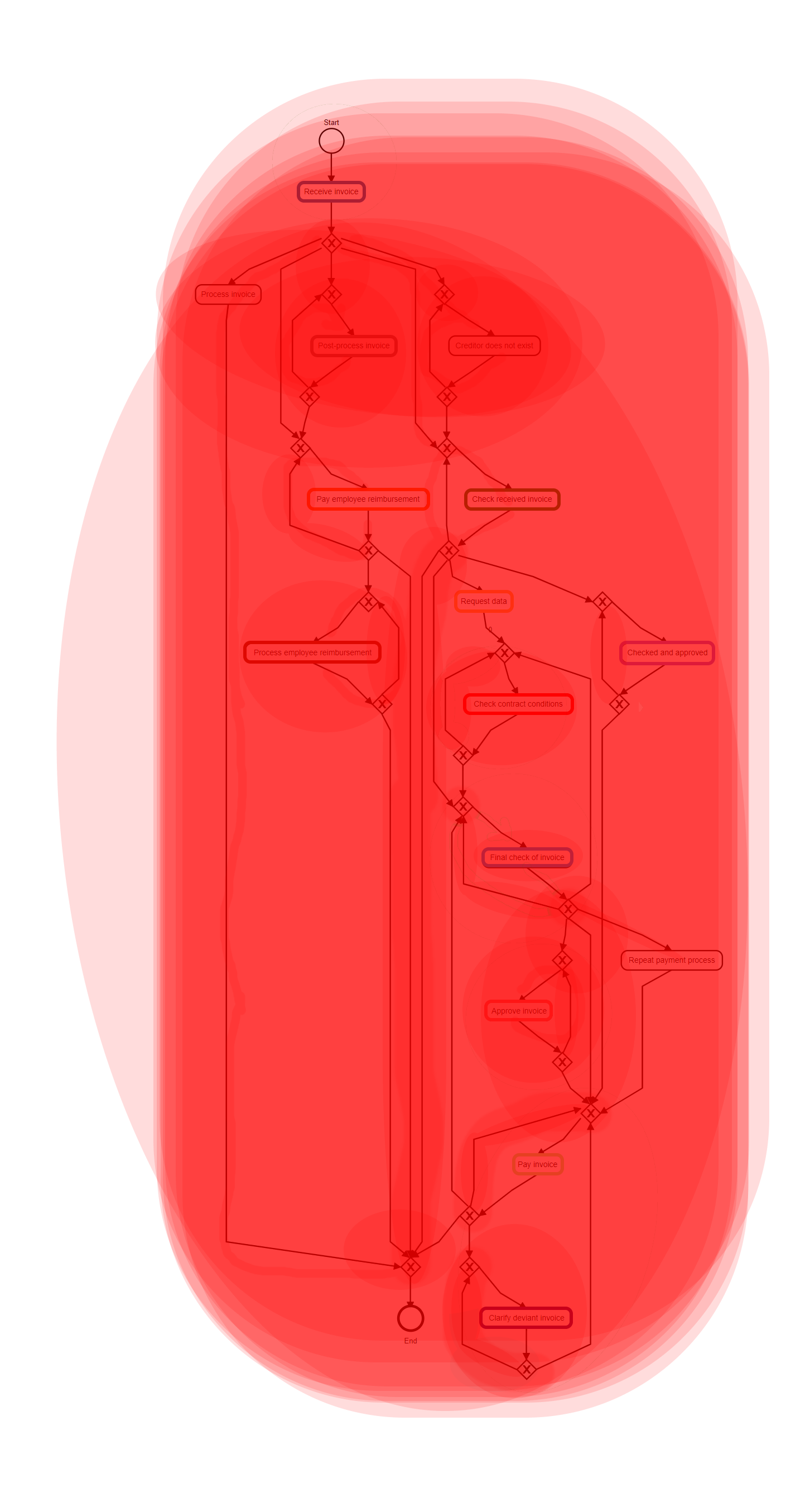}
    }
    \par
    \subfloat[PIM\textsubscript{30}]{
        \includegraphics[width=.12\linewidth]{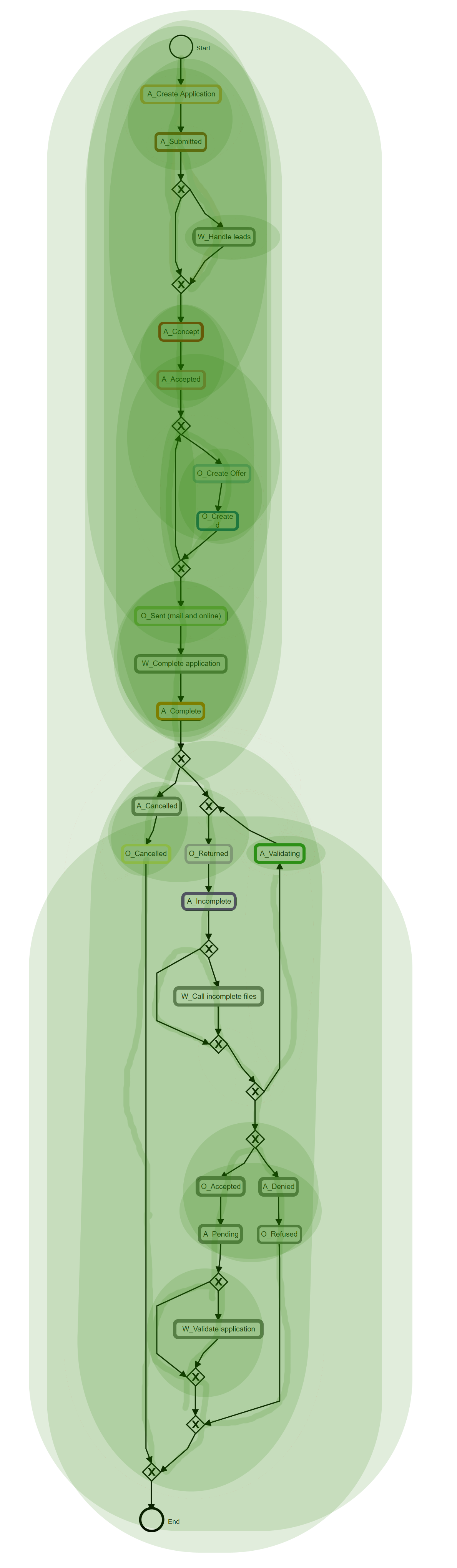}
        \includegraphics[width=.08\linewidth]{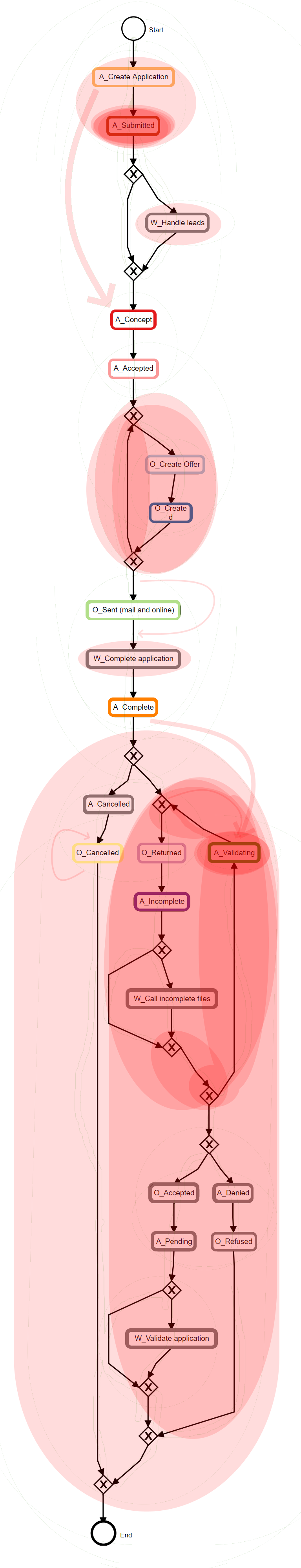}}
    \subfloat[IMf]{
        \includegraphics[width=.15\linewidth]{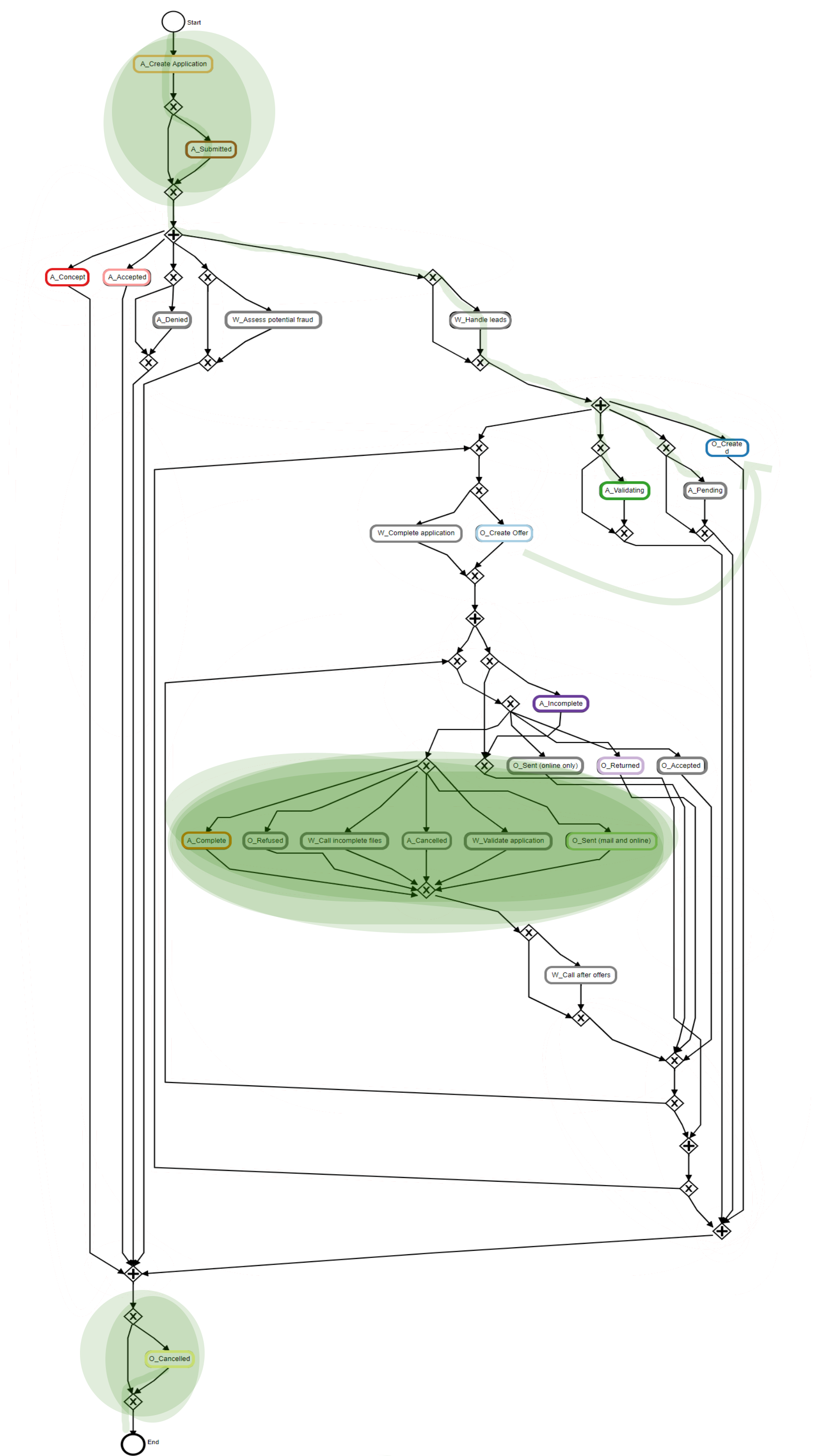}
        \includegraphics[width=.15\linewidth]{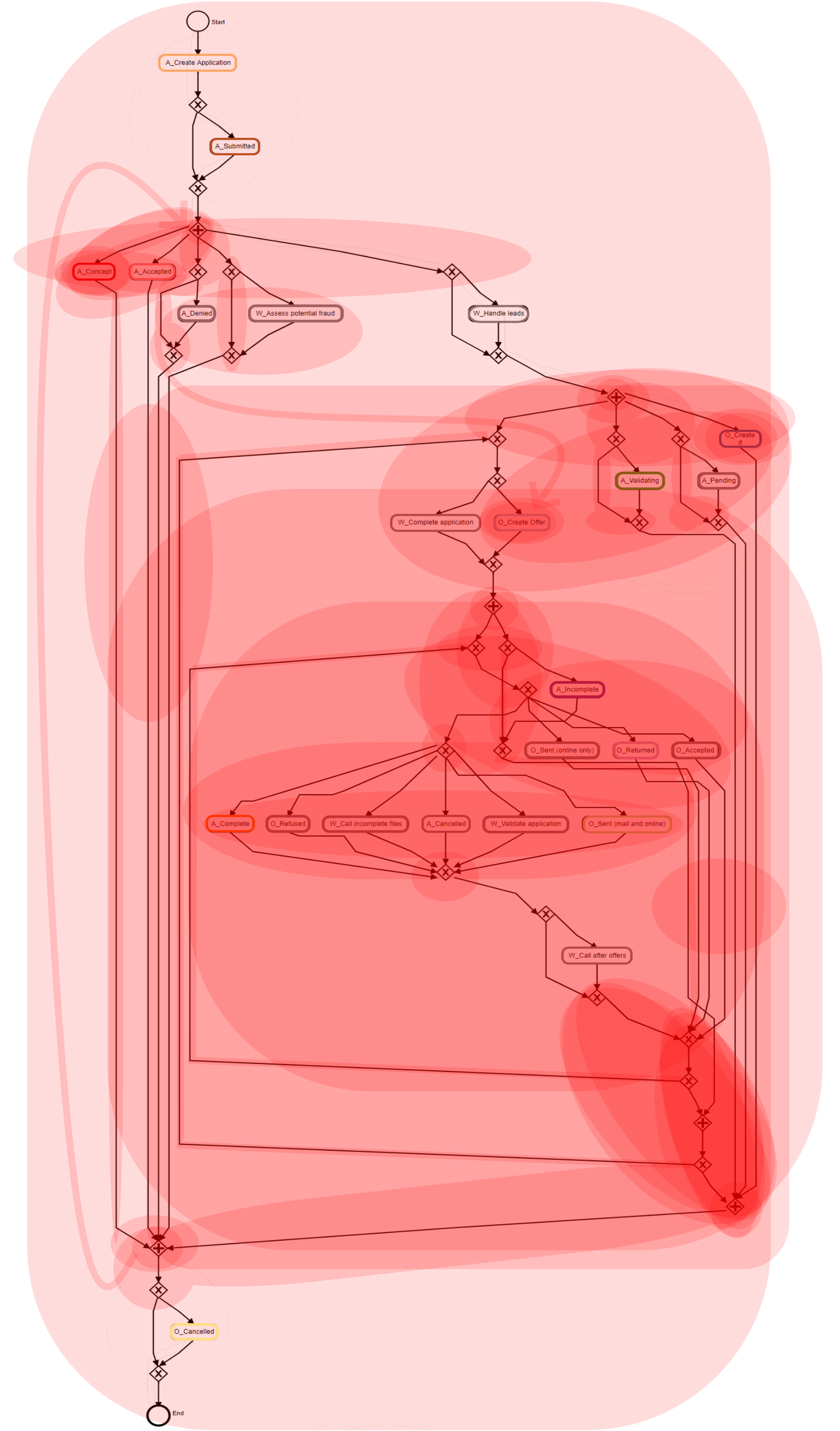}
    }
    \subfloat[SM]{
        \includegraphics[width=.15\linewidth]{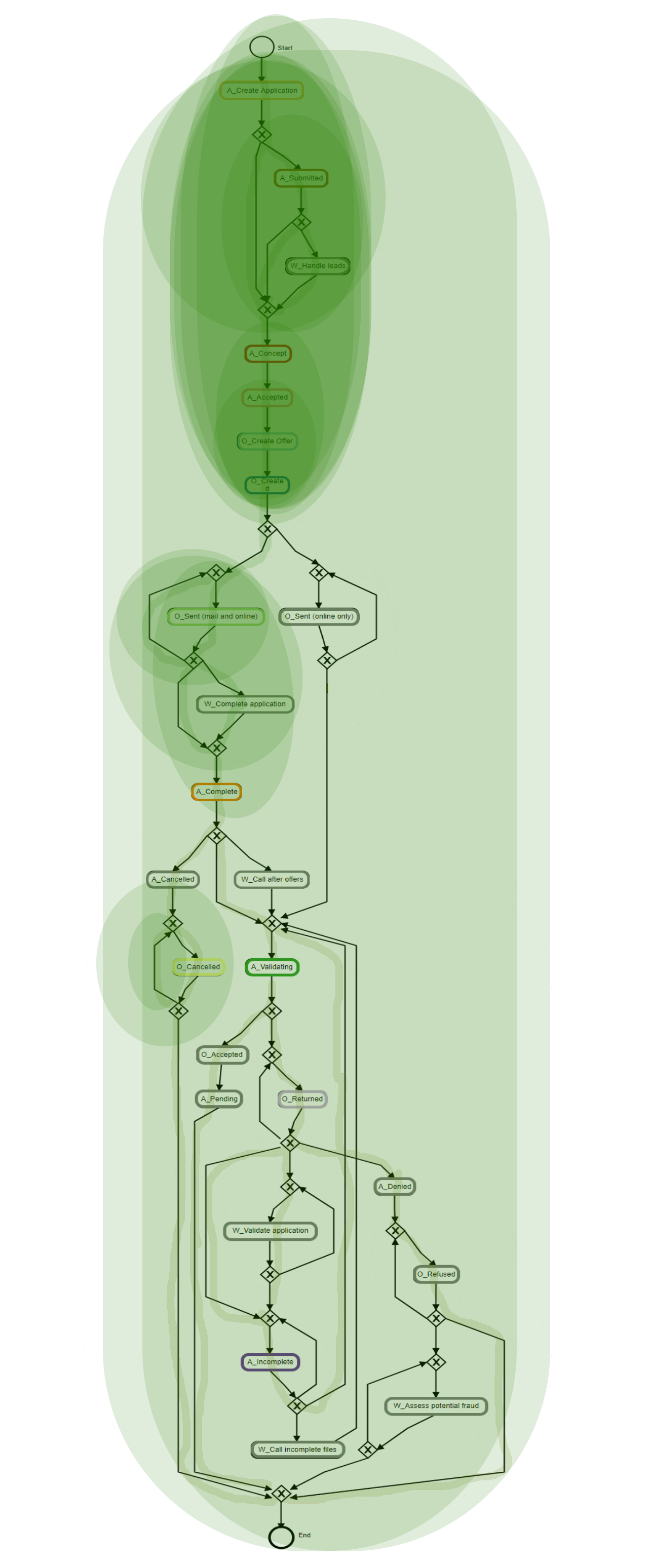}
        \includegraphics[width=.15\linewidth]{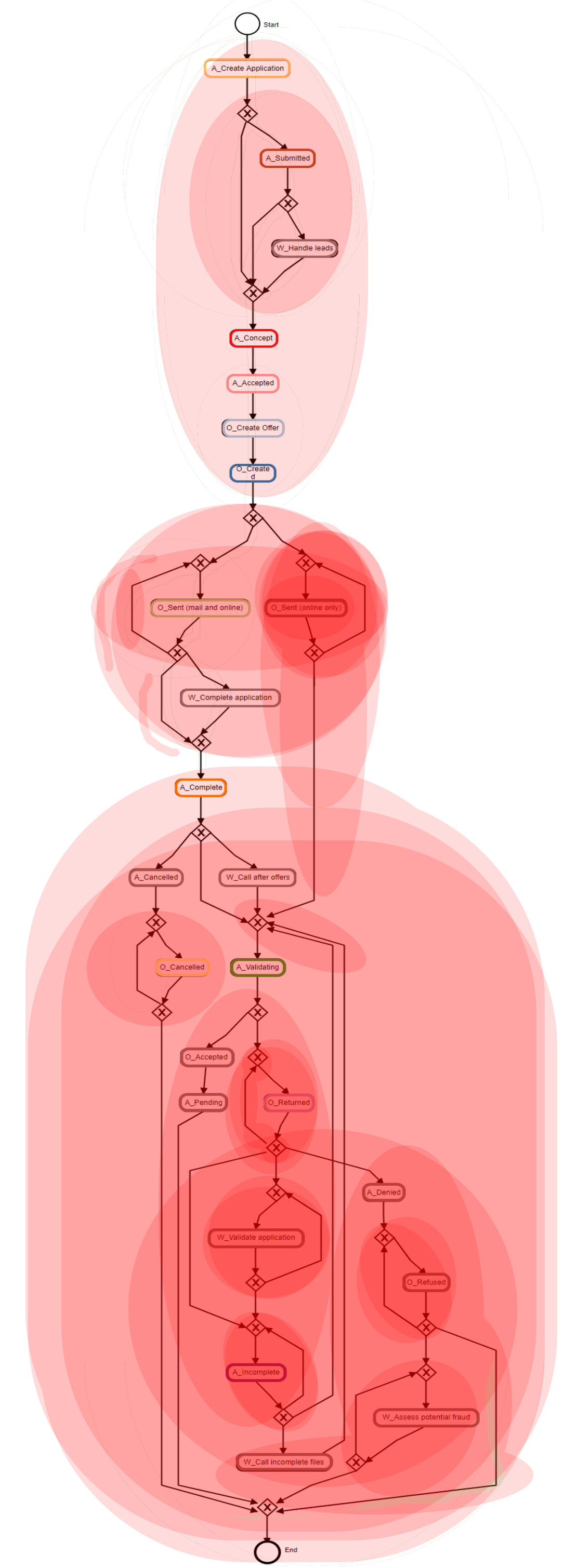}
    }
    \caption{Heatmaps of ``good'' and ``bad'' model structures for the RTFMP event log (a)-(c), Invoice event log (d)-(f), BPIC17\textsubscript{cp} event log (g)-(i).}
    \label{fig:heatmap:all}
\end{figure}

Detailed results on the benchmark event logs are given in Table~\ref{tab-app:results}. Detailed results on the additional event logs are given in Table~\ref{tab-app:part2results}.
Figure~\ref{fig:heatmap:all} shows the heat maps of user annotations regarding what users considered ``good'' (understood and trusted) and ``bad'' (too complex, not trusted) model structures for the RFTMP, Invoice, and BPIC17\textsubscript{cp} log.

\begin{table}[!h]

\footnotesize
\begin{tabular}{ | @{}C{3.5em} || @{}C{2.1em} | @{}C{1.8em} | @{}C{1.8em} | @{}C{1.6em} | @{}C{1.2em} | @{}C{1.5em} | @{}C{1.5em} | @{}C{1.2em} | @{}C{1.2em} | @{}C{4.5em} |}
\rot{75}{\textit{Log}} & \rot{75}{\textit{Method}} & \rot{75}{\textit{Fitness}} & \rot{75}{\textit{Precision}} & \rot{75}{\textit{F-score}} &  \rot{75}{\textit{Size}} & \rot{75}{\textit{Edges}} & \rot{75}{\textit{CFC}} & \rot{75}{\textit{Structured}} &\rot{75}{\textit{Sound}} & \rot{75}{\textit{Time(s)}} \\
\hline
\hline
\multirow{3}{*}{BPIC14} & IM & .90 & .61  & \textbf{.73} &  76 & 120 & 56 & \textbf{yes} & \textbf{yes} & 8.27  \\
& SM & \textbf{.97} & .48 & .65 & 100 & 266 & 194 & no & no & \textbf{ .78}  \\
& PIM$_{30}$ & .52 & \textbf{.64} & .57 &  \textbf{32} & \textbf{43} & 17 & \textbf{yes} & \textbf{yes} & 74.64\\
& PIM & .57 & .45 & .5 & 66 & 92 & \textbf{11} & \textbf{yes} & \textbf{yes} & \textbf{.94}\\
\hline
\multirow{3}{*}{BPIC17\textsubscript{LC}} & IM & .85 & .15 & .25 &  168 & 247 & 103 & \textbf{yes} & \textbf{yes} & 22.64  \\
& SM & \textbf{.96} & .73 & .83 & 139 & 201 & 100 & no  & no & \textbf{1.18}  \\
& PIM$_{30}$ & .68 & \textbf{.95} & .79 & \textbf{52} & \textbf{61} & \textbf{17} & \textbf{yes} & \textbf{yes} & 9,757\\
& PIM & .86 & .91 & \textbf{.88} & 81 & 102 & 36 & \textbf{yes} & \textbf{yes} & 3.2\\
\hline
\multirow{3}{*}{BPIC17\textsubscript{cp}} & IM & t/o & t/o & t/o &  60 & 87 & 37 & \textbf{yes} & \textbf{yes} & 4.82  \\
& SM & \textbf{.92} & .986 & \textbf{.95} & 51 & 69 & 21 & no  & \textbf{yes} & 0.41  \\
& PIM$_{30}$ & .86 & \textbf{.999} & .92 & \textbf{37} & \textbf{43} & \textbf{14} & \textbf{yes} & \textbf{yes} & 21.1\\
& PIM & .90 & .98 & .94 & \textbf{37} & \textbf{45} & \textbf{14} & \textbf{yes} & \textbf{yes} & \textbf{0.2}\\
\hline
\multirow{3}{*}{Invoice} & IM & .88  & .76 & .82  & 34 & 44 & 18 & \textbf{yes} & \textbf{yes} & .98  \\
& SM & \textbf{1.00}  & .92 & \textbf{.96} &  41 & 64 & 36 & no & \textbf{yes} & .25  \\
& PIM$_{30}$ & .88 & \textbf{.96} & .92 & \textbf{17} & \textbf{19} & \textbf{6} & \textbf{yes} & \textbf{yes} & \textbf{.01}\\
& PIM & .61 & .94 & .73 & 19 & 23 & 8 & \textbf{yes} & \textbf{yes} & .01\\
\hline
\multirow{3}{*}{P2P} & IM & .80 & .77 & .78 & 101 & 134 & 70 & \textbf{yes} & \textbf{yes} & 164.07  \\
& SM & \textbf{1.00} & .61 & .76 & 114 & 287 & 207 & no & no & 8.89  \\
& PIM$_{30}$ & .69 & \textbf{.94}  & \textbf{.80} & \textbf{35} & \textbf{45} & \textbf{19} & \textbf{yes} & \textbf{yes} & \textbf{3.09}\\
& PIM & .74 & .83 & .78 & 37 & 50 & 20 & \textbf{yes} & \textbf{yes} & 3.1\\
\hline
\end{tabular}
\caption{Evaluation results of IMf, SM, PIM$_{30}$ and PIM on 4 additional event logs; \textbf{bold} = best value per log.}
\label{tab-app:part2results}
\end{table}


\clearpage
\section{Example of Visual Alignments}\label{sec-app:visual-alignment}

\begin{figure}[!h]
    \centering
    \includegraphics[width=\linewidth]{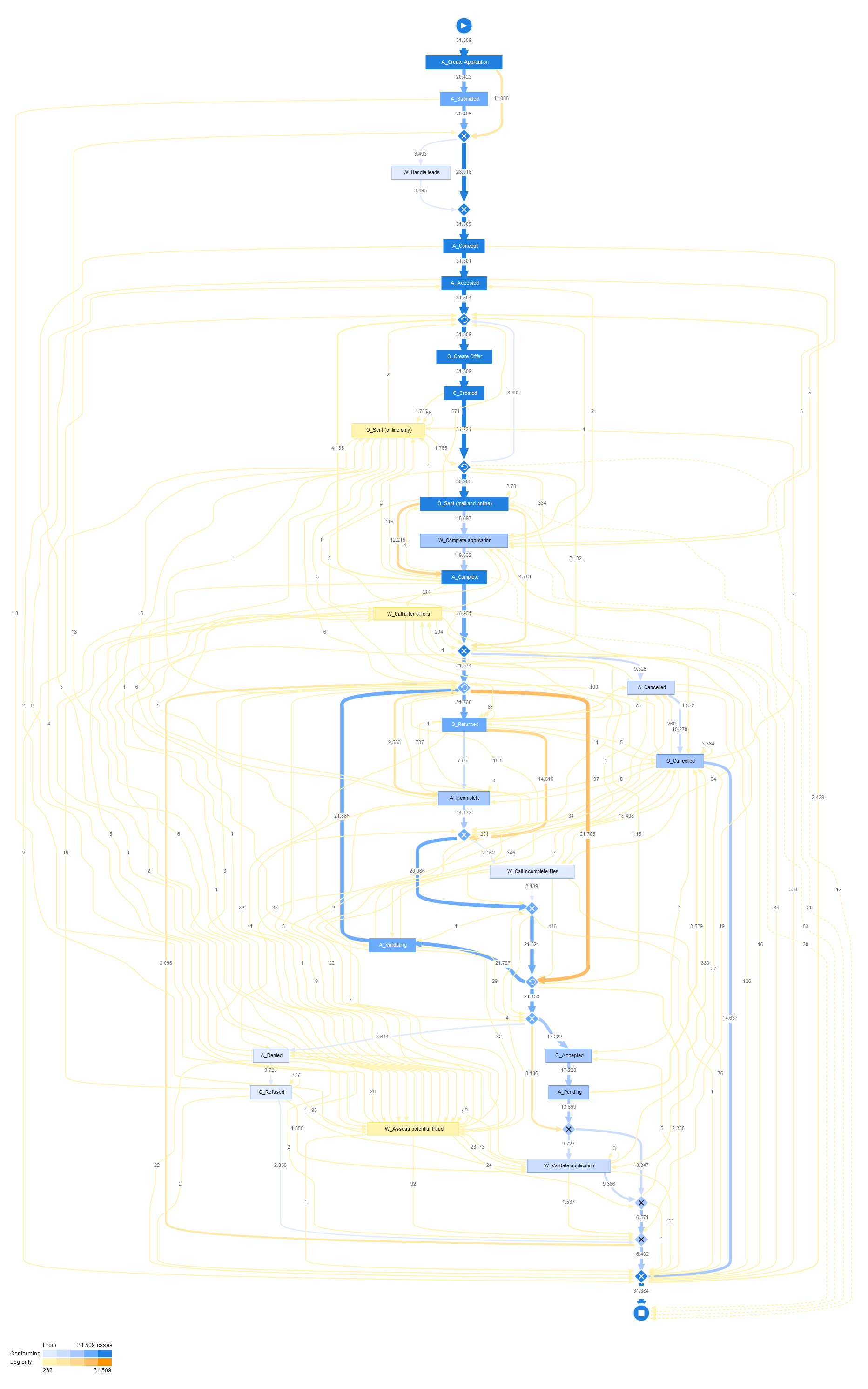}
    \caption{Visual Alignments of BPIC17\textsubscript{cp} on the PIM model in UiPath Process Mining}
    \label{fig:visual_alignment}
\end{figure}

Fig.~\ref{fig:visual_alignment} shows an example of a complete visual alignment~\cite{Bie2019VisualThesis} of BPIC17\textsubscript{cp} on the PIM model in the UiPath Process Mining platform; blue nodes and edges describe the discovered model; yellow edges are the visual alignment showing the behavior in the event log deviating from the discovered model.

\end{appendices}
\end{document}